\documentclass[twocolumn,twocolappendix]{aastex7}

\usepackage{upgreek}
\usepackage{amsmath}
\usepackage{booktabs}
\usepackage{tabularx}
\usepackage{adjustbox}

\received{}
\revised{}
\accepted{}

\submitjournal{ApJ}

\shorttitle{Sonora Flame Skimmer}
\shortauthors{Mang et al.}

\graphicspath{{./}{figures/}}

\begin{document}

\title{The Sonora Substellar Atmosphere Models. VII. Flame Skimmer: Cloud-free Atmospheric and Evolutionary Models for the Coldest Substellar Objects}

\correspondingauthor{James Mang}
\email{j\_mang@utexas.edu}

\author[0000-0001-5864-9599]{James Mang}
\altaffiliation{NSF Graduate Research Fellow.}
\affiliation{Department of Astronomy, University of Texas at Austin, Austin, TX 78712, USA}
\email{j\_mang@utexas.edu}

\author[0000-0003-1728-8269]{Yayaati Chachan}
\affiliation{Department of Astronomy \& Astrophysics, University of California Santa Cruz, Santa Cruz, CA 95064, USA}
\email{ychachan@ucsc.edu}

\author[0000-0002-4404-0456]{Caroline V. Morley}
\affiliation{Department of Astronomy, University of Texas at Austin, Austin, TX 78712, USA}
\email{cmorley@utexas.edu}

\author[0000-0003-1240-6844]{Natasha E. Batalha}
\affiliation{Space Science and Astrobiology Division, NASA Ames Research Center, Moffett Field, CA, 94035, USA}
\email{natasha.e.batalha@nasa.gov}

\author[0000-0002-0413-3308]{Nicholas F. Wogan}
\affiliation{NASA Ames Research Center, Moffett Field, CA 94035}
\affiliation{SETI Institute, Mountain View, CA 94043}
\email{nicholas.f.wogan@nasa.gov}

\author[0000-0003-1622-1302]{Sagnick Mukherjee}
\altaffiliation{51 Pegasi b Fellow}
\affiliation{School of Earth and Space Exploration, Arizona State University, Tempe, AZ, USA}
\email{smukhe50@asu.edu}

\author[0000-0002-9843-4354]{Jonathan J. Fortney}
\affiliation{Department of Astronomy \& Astrophysics, University of California Santa Cruz, Santa Cruz, CA 95064, USA}
\email{jfortney@ucsc.edu}

\author[0000-0002-5251-2943]{Mark S. Marley}
\affiliation{Department of Lunar and Planetary Sciences, University of Arizona, Tucson, AZ 85721, USA}
\email{marksmarley@arizona.edu}

\author[0000-0001-6627-6067]{Channon Visscher}
\affiliation{Chemistry \& Planetary Sciences, Dordt University, Sioux Center, IA 51250, USA}
\affiliation{Center for Extrasolar Planetary Systems, Space Science Institute, Boulder, CO 80301, USA}
\email{Channon.Visscher@dordt.edu}

\author[0000-0002-4088-7262]{Ehsan Gharib-Nezhad}
\affiliation{Space Science and Astrobiology Division, NASA Ames Research Center, Moffett Field, CA, 94035, USA}
\email{}

\begin{abstract}
JWST has provided unprecedented access to ultra-cool brown dwarfs and has pushed the boundaries of directly imaging temperate giant planets. As we continue to push toward detecting sub-Saturn and Neptune-like planets, it is crucial to develop atmospheric and evolutionary models that better capture the complexity and diversity of planetary atmospheres similar to the gas and ice giants in our Solar System. We present \texttt{Sonora Flame Skimmer}, the next suite of cloud-free 1D atmospheric and evolutionary models in chemical equilibrium and disequilibrium probing colder temperatures (down to 50 K), smaller objects (down to log(g) = 2), and a wide range of metallicities (10x sub-solar to 100x super-solar) and C/O ratios (solar to 2.5x solar). Beyond expanding the physical parameter space of previous Sonora models, we update the opacities and evolutionary model framework from \texttt{Sonora Bobcat}, as well as the chemical treatment of volatiles (H$_2$O, CH$_4$, NH$_3$) and carbon species such as CO$_2$ from \texttt{Sonora Elf Owl}. For the evolution of these substellar objects, we find that high-metallicity atmospheres lead to slower cooling compared to solar metallicity, while the strength of vertical mixing ($K_{\rm zz}$) has a negligible impact on the evolutionary tracks. At the highest metallicity explored here (100$\times$ solar), the deuterium-burning and hydrogen-burning minimum masses fall to 5.39 and 45.03 $M_{\rm J}$, respectively. All the models presented here, including the atmospheric structure, chemical profiles, spectra, synthetic photometry, and evolutionary models, are publicly available.
\end{abstract}



\section{Introduction} \label{sec:intro}

Over the past two decades, atmospheric models have been extensively used to interpret the observations of directly imaged exoplanets and brown dwarfs \citep[e.g.][]{Marley1996, Allard1996, Marois2008, Lagrange2009, Macintosh2015}, providing key insights into the chemistry and dynamics of extrasolar worlds. Until recently, we could only directly image planets that were much warmer, brighter, and younger than the planets in our solar system; we lacked the capability to directly image and characterize the atmospheres of mature temperate giant planets analogous to the gas giants in our own Solar System. While the discovery of such planets through radial velocity has provided essential constraints on orbital architectures and dynamical masses, without directly imaging them, we are unable to assess how representative Jupiter and Saturn's atmospheres and compositions are of other mature giant planets. 

This has positioned brown dwarfs as key atmospheric laboratories, spanning similar effective temperatures and masses as these cool giant planets \citep{cushing2011}. Y dwarfs in particular ($T_{\rm eff} < 500$~K) have provided critical insight into the chemistry and dynamics of ultra-cool atmospheres. The coldest Y-dwarf, WISE J085510.83-70071442.5 \citep[][Miles et al. submitted, Kothari et al. submitted]{Luhman2014, skemer2016, Luhman2024, Rowland2024, Kuhnle2025} represents the closest brown dwarf to bridge the temperature regime toward true-Jupiter analogs.



\subsection{Ongoing JWST Observations of Cold Worlds}

With JWST, we are rapidly expanding the sample of cold substellar objects that we can observe at moderate resolution. JWST has already obtained spectra of dozens of Y dwarfs spanning effective temperatures of 250–450 K \citep[e.g.][]{Beiler2023, Luhman2024, Lew2024, Beiler2024sample, Faherty2024, Rowland2024, Suarez2025, Kuhnle2025, DeFurio2025}. Its broad wavelength coverage (0.6–28 $\upmu$m) and moderate spectral resolution have transformed our ability to probe cold atmospheres. Observations of Y-type brown dwarfs have revealed atmospheric properties that deviate from expectations based on chemical equilibrium, including enhanced CO$_2$ abundances \citep{Beiler2024}, CH$_4$ emission features \citep{Faherty2024}, and low PH$_3$ abundances \citep{Beiler2024, Rowland2024, Burgasser2025}. So far, these programs have demonstrated that disequilibrium chemistry is ubiquitous in substellar atmospheres along with clear signs of variability \citep[Miles et al. submitted]{Biller2024, McCarthy2025, Nasedkin2025, Oliveros-Gomez2026}.

Beyond isolated brown dwarfs, JWST is the first facility capable of directly imaging RV-detected temperate giant planets such as $\epsilon$~Indi Ab \citep{Matthews2024, Matthews2026, Sanghi2026epsindi} and 14 Her c \citep{BardalezGagliuffi2025}. Other programs targeting true-Jupiter analogs like $\epsilon$~Eri b \citep{Sanghi2026epseri} have thus far eluded clear detections but provide essential limits for atmospheric scenarios that are viable for future observations. JWST is also beginning to discover a new population of lower-mass, colder directly imaged planets, such as TWA-7 b \citep{Lagrange2025, Crotts2025}, a young sub-Saturn. These discoveries represent the first steps toward directly imaging true Solar System analogs with many other programs pushing the limits towards even smaller objects. (GO 4050, GO 5835, GO 10764 PI: Carter, GO 6005 PI: Biller, GO 6122, 8581 PI: Bowens-Rubin)

\subsection{A Need for Colder Atmosphere and Evolution Models}

Existing atmospheric model grids have provided a broad foundation for interpreting substellar objects. These models fall into several categories that differ in their treatment of clouds, chemistry, coupling to interior evolution, and exploration of parameter space. Early models such as \citet{Burrows1997} and \texttt{COND} \citep{Baraffe2003} focused on cloud-free atmospheres and equilibrium chemistry. Next, the coupled atmospheric and evolutionary framework of \citet{Saumon2008} linked radiative–convective equilibrium atmospheres with hybrid models including clouds.

The Sonora family of models builds on the legacy of \citet{Saumon2008} by extending this coupled atmosphere–evolution framework with updated opacities, chemistry, and cloud treatments. \texttt{Sonora Bobcat} \citep{Marley2021} established a baseline grid of clear, equilibrium-chemistry atmospheric and evolutionary models. \texttt{Sonora Cholla} \citep{Karalidi2021} and \texttt{Sonora Elf Owl} \citep{Mukherjee2024, Wogan2025} extended this framework to include disequilibrium chemistry in the atmospheres, while \texttt{Sonora Diamondback} \citep{Morley2024} incorporated refractory cloud species (e.g., silicates and iron) for warmer L and T dwarfs and explored their impact on both atmospheric structure and evolution. \texttt{Sonora Red Diamondback} \citep{Davis2025} then extended the \texttt{Sonora Diamondback} evolutionary models to higher effective temperatures by incorporating SPHINX atmospheric models as boundary conditions. A summary of the full parameter space explored across the Sonora family of models is provided in Table~\ref{table:parameters}.

Models from other groups have also explored complementary approaches. For example, BT-Settl \citep{Allard2011} and BHAC-15 \citep{Baraffe2015} also include cloud treatments. The Exo-REM framework \citep{Charnay2018} provides flexible forward models that include parameterized cloud treatments and disequilibrium chemistry, particularly suited for directly imaged exoplanets. The ATMO2020 \citep{Phillips2020} and \citet{Tremblin2015, Tremblin2019} models explore both equilibrium and disequilibrium chemistry and alternative non-adiabatic thermal structures. The SAND models \citep{Alvarado2024} characterize hotter ($T_{\rm eff} \geq$ 700~K), ultra-low metallicity objects that are being found in the halo and thick disk and have atmospheric and evolutionary models with disequilibrium chemistry. The \textit{coolTLUSTY} models \citep{Lacy2023}, extend into the Y dwarf regime with water clouds and disequilibrium chemistry. On the evolutionary side, the BEX models \citep{Linder2019} provide tracks for low-mass objects spanning 0.16–2 M$_{\rm Jup}$ using atmospheric models with equilibrium chemistry and no water clouds. Unlike most model grids that self-consistently couple a single atmospheric framework to the evolutionary calculations, the BEX models combine atmospheric boundary conditions from multiple model families and codes (e.g., \texttt{COND} \citep{Allard2001}, \texttt{HELIOS} \citep{Malik2017}, and \texttt{petitCODE} \citep{Molliere2015, Molliere2017}), relying on interpolation and extrapolation when extending beyond the parameter space covered by the underlying atmospheric grids.

Despite this progress, existing grids of models provide mostly overlapping coverage of parameter space (e.g. $T_{\rm eff}$ and log(g)), but interpreting the coldest substellar atmospheres now accessible with JWST remains challenging. In particular, expanding self-consistent atmospheric and evolutionary models to include disequilibrium chemistry and volatile condensation across a broader range of metallicities, surface gravities, and effective temperatures approaching those of Neptune- and Uranus-like planets will further enhance our ability to discover and interpret the observations of these objects. No existing Sonora grid extends to temperatures below $\sim$200 K, metallicities as high as 100$\times$ solar, or surface gravities as low as $\log(g)=2$ (cgs; Table~\ref{table:parameters}). The evolutionary models within the Sonora family have also only been computed for grids in chemical equilibrium as \texttt{Sonora Cholla} and \texttt{Sonora Elf Owl} do not have evolution models accompanying their atmospheric models. This motivated the development of \texttt{Sonora Flame Skimmer}.

\begin{table*}
\centering
\setlength{\tabcolsep}{3pt}

\begin{adjustbox}{width=1.12\textwidth, right}
\begin{tabular}{|c|c|c|c|c|c|}
\hline
\textbf{Parameter} & \textbf{Flame Skimmer} & \textbf{Bobcat} & \textbf{Cholla} & \textbf{Diamondback} & \textbf{Elf Owl} \\
\hline

Reference & this work & \citet{Marley2021} & \citet{Karalidi2021} & \citet{Morley2024} & \hspace{-20pt}\begin{tabular}{c} \citet{Mukherjee2024} \\ \citet{Wogan2025} \end{tabular} \\
\hline

$T_{\mathrm{eff}}$ [K] (step) & 50--2400 (25--100) & 200--2400 (25--100) & 500--1300 (50--100) & 900--2400 (100) & 275--2400 (25--100) \\
\hline

$\log(g)$ [cgs] (step) & 2.0--5.5 (0.25) & 3.25--5.5 (0.25--0.5) & 3.0--5.5 (0.25) & 3.5--5.5 (0.5) & 3.25--5.5 (0.25) \\
\hline

[M/H] [dex] &
\hspace{-20pt}\begin{tabular}{c}
$-$1.0, $-$0.5, 0.0, +0.5, \\
+1.0, +1.5, +2.0
\end{tabular}
& $-$0.5, 0.0, +0.5
& 0.0
& $-$0.5, 0.0, +0.5
& \hspace{-20pt}\begin{tabular}{c}
$-$1.0, $-$0.5, 0.0, \\
+0.5, +0.7, +1.0
\end{tabular} \\
\hline

$f_{\mathrm{sed}}$ & \dots & \dots & \dots & 1, 2, 3, 4, 8, nc & \dots \\
\hline

$\log(K_{\rm zz}$) [cgs] & EQ \& 2, 4, 7, 8, 9 & EQ & 2, 4, 7, 10 & EQ & 2, 4, 7, 8, 9 \\
\hline

C/O ratio & 0.229, 0.458, 0.687, 1.14 & 0.229, 0.458, 0.687 & 0.458 & 0.458 (solar) & 0.229, 0.458, 0.687, 1.14 \\
\hline

Evolution Models & Yes & Yes & No & Yes & No \\
\hline
\end{tabular}
\end{adjustbox}
\caption{Grid Parameters for the Sonora Family of Models}
\label{table:parameters}
\end{table*}

\subsection{Sonora Flame Skimmer: Bridging Solar System-analogs and Brown Dwarfs}

In this paper, we present \texttt{Sonora Flame Skimmer}, the newest set of atmospheric and evolutionary models in the Sonora family. The name “Flame Skimmer” refers to the bright orange dragonfly \textit{Libellula saturata}, commonly known as the firecracker skimmer and native to the Sonoran Desert. The \texttt{Sonora Flame Skimmer} models span 45 effective temperatures from 50 to 2400 K and 15 surface gravities of $\log(g) = 2.0$–5.5, with 7 metallicities [M/H] = $[-1.0, -0.5, 0.0, +0.5, +1.0, +1.5, +2.0]$ (scaled relative to solar; \citealt{lodders2009}) and 4 C/O ratios of [0.229, 0.458, 0.687, and 1.14] where the solar C/O ratio is 0.458 \citep{lodders2009}. For models including disequilibrium chemistry, we adopt eddy diffusion coefficients of $K_{\rm zz} = 10^2, 10^4, 10^7, 10^8$, and $10^9$ cm$^2$ s$^{-1}$ for consistency with the values adopted in the \texttt{Sonora Elf Owl} grid.  The \texttt{Sonora Flame Skimmer} grid comprises 94,500 atmospheric models computed in chemical disequilibrium and 18,900 models in chemical equilibrium. The full parameter space of the grid is summarized in Table \ref{table:parameters}, alongside the rest of the Sonora family of models. 

Overall, \texttt{Sonora Flame Skimmer} expands the parameter space of all the previous Sonora grids to colder temperatures, lower surface gravities, and more extreme metallicities. These models will enable the direct characterization of Neptune-like objects that we could not have done before within the Sonora family of models. The grid also incorporates updated opacity lists (Section \ref{sec:picaso}) and revised evolutionary model assumptions (Section \ref{sec:evolution}). Several key differences relative to earlier Sonora model families are summarized below:

\begin{enumerate}
    \item For equilibrium chemistry models, \texttt{Sonora Flame Skimmer} adopts the updated opacities introduced in the 1460 grid, superseding those used in the 1060 grid (\texttt{Sonora Bobcat}; see Section \ref{sec:picaso}). We also expand the grid to much larger ranges of metallicity and lower surface gravities.

    \item In \texttt{Sonora Elf Owl}, the abundances of quenched species were held fixed above the quench level. In contrast, in \texttt{Sonora Flame Skimmer} we allow H$_2$O, NH$_3$, and CH$_4$ to condense once they reach their saturation vapor pressures in disequilibrium models. This accounts for the depletion of these volatiles from the gas-phase atmospheric composition and chemistry, but does not include cloud opacity. This treatment becomes particularly important for the coldest objects in the grid ($T_{\rm eff} \leq 275$~K), where these species are expected to condense out of the gas phase; see Section~\ref{sec:chemistry}. We also update the metallicity dependence of the chemical timescales used to determine the quench points.
    
    \item \texttt{Sonora Elf Owl} and \texttt{Cholla} provided atmospheric models without accompanying evolutionary tracks. Here, we compute evolutionary tracks for models with disequilibrium chemistry for the first time. Relative to \texttt{Sonora Bobcat}, the new evolutionary models also adopt an updated equation of state, along with additional changes described in Section \ref{sec:evolution}.
\end{enumerate}

Here we present the details of \texttt{Sonora Flame Skimmer}. In Section~\ref{sec:methods}, we describe the framework used to generate the atmospheric models (Section~\ref{sec:picaso}), including updates to the chemistry and opacities (Section~\ref{sec:chemistry}), as well as the evolutionary modeling approach (Section~\ref{sec:evolution}). In Section~\ref{sec:results}, we present the key results from the grid, including the thermal structures (Section~\ref{sec:ptprofiles}), spectra (Section~\ref{sec:spectra}), evolutionary tracks (Section~\ref{sec:evoresults}), and color–magnitude diagrams (Section~\ref{sec:cmd}). In Section~\ref{sec:discussion}, we discuss convergence behavior in specific regions of the grid, the role of clouds in future model developments, and applications of the grid to JWST and future missions. We summarize our findings in Section~\ref{sec:conclusion}. The thermal structures, spectra, and evolutionary tracks from this grid are publicly available on Zenodo\footnote{\href{https://doi.org/10.5281/zenodo.20030439}{Flame Skimmer Zenodo Repository}}. 

\section{Models} \label{sec:methods}

\subsection{Atmospheric Modeling with \texttt{PICASO}} \label{sec:picaso}
We use \texttt{PICASO} \citep{Mukherjee2023, Mang2026}, an open-source Python-based atmospheric model to compute one-dimensional pressure--temperature (P-T) profiles in RCE and is rooted in the legacy of the substellar \texttt{EGP} code \citep{Marley1996, MarleyMcKay1999, Fortney2005, Fortney2007, Fortney2008, Robinson2014, Morley2018, Marley2021, Karalidi2021, Morley2024}. \texttt{PICASO} has been used to model exoplanet and brown dwarf atmospheres in numerous JWST programs \citep[e.g.][]{Beiler2023, Miles2023, Rustamkulov2023, Biller2024, Lew2024, Alderson2024, Crotts2025, BardalezGagliuffi2025, Beichman2025, Mukherjee2025gj436b, Matthews2026, Sanghi2026epsindi, Sanghi2026epseri} and was used to generate the \texttt{Sonora Elf Owl} grid.

\begin{table}
    \centering
    \begin{tabular}{
    c|p{0.8\columnwidth}}
        Molecule & Reference for Opacities \\
        \hline
         C$_2$H$_2$ & \citet{hitran2012}\\
         C$_2$H$_4$ & \citet{hitran2012}\\
         C$_2$H$_6$ & \citet{hitran2012} \\
         CH$_4$ & \citet{yurchenko13vibrational}, \citet{yurchenko2014}, \citet{STDS}, \citet{Pine:1992}, \\
            & \citet{Hargreaves2020ch4}, computed with methods in 
         \citet{GharibNezhad2021}\\
         CO &  \citet{HITEMP2010,HITRAN2016,li15rovibrational}\\
         CO$_2$ &  \citet{HUANG2014reliable}\\
         CrH &  \citet{Burrows02CrH}, computed in \citet{GharibNezhad2021}\\
         Cs & \citet{Ryabchikova2015} \\
         Fe &  \citet{Ryabchikova2015,oBrian1991Fe,Fuhr1988Fe, Bard1991Fe,Bard1994Fe} \\
         FeH &  \citet{Dulick2003FeH}, with added E-A transition \citet{Hargreaves2010FeH}  \\
         H$_2$ & \citet{HITRAN2016} \\
         H$_3^+$ &  \citet{Mizus2017H3p}\\
         H$_2$--H$_2$ & \citet{Saumon12} with added overtone from \citet{Lenzuni1991h2h2} Table 8\\
         H$_2$--He &  \citet{Saumon12} \\
         H$_2$--N$_2$ &  \citet{Saumon12} \\
         H$_2$--CH$_4$ &  \citet{Saumon12} \\
         H$_2^-$  &  \citet{bell1980free}\\
         H$^-$ bf &  \citet{John1988H}\\
         H$^-$ ff &  \citet{Bell1987Hff}\\
         H$_2$O &  \citet{Polyansky2018H2O}\\
         H$_2$S &  \citet{azzam16exomol}\\
         HCN &  \citet{Harris2006hcn,Barber2014HCN,hitran2020}\\
         LiCl &  \citet{Bittner2018Lis}\\
         LiF &  \citet{Bittner2018Lis}\\
         LiH & \citet{Coppola2011LiH} \\
         MgH & \citet{Yadin2012MgH,GharibNezhad2013MgH} computed in \citet{GharibNezhad2021}\\
         N$_2$ &  \citet{hitran2012}\\
         NH$_3$ &  \citet{Coles2019NH3,ExoMol15NH3}, computed with methods in \citet{exomolop} \\
         OCS &  \citet{HITRAN2016}\\
         PH$_3$ & \citet{sousa14exomol} \\
         Rb & \citet{Ryabchikova2015} \\
         SiO &  \citet{Barton2013SiO, GharibNezhad2021} \\
         SO$_2$ &  \citet{underwood2016exomol} \\
         TiO & \citet{McKemmish2019TiO} computed in \citet{GharibNezhad2021}\\
         VO &   \citet{McKemmish16} computed in \citet{GharibNezhad2021}\\
         Li, Na, K &  \citet{Ryabchikova2015,Allard2007AA, Allard2007EPJD,Allard2016, Allard2019}, as compiled in \citet{Molliere2019} \\
         
    \end{tabular}
    \caption{References of gaseous opacities used for calculating the resulting spectra in this work. If not specified, the cross section was computed and compiled with the framework discussed in \citet{Freedman2008,Freedman2014}.}
    \label{tab:opatab}
\end{table}  

Each 1D atmosphere is discretized into 90 pressure layers (91 grid levels). The pressure grid for each model is adopted from the \texttt{Sonora Elf Owl} grid. For models colder than 275 K (the lower temperature limit of \texttt{Sonora Elf Owl}), we adopt the same pressure grid as the 275 K model.

Gas opacities are computed using the correlated-$k$ method. The opacity database is based on \citet{Freedman2008}, with updates from \citet{Freedman2014}, and includes revised opacities for H$_2$O, FeH, and alkali metals, as well as the addition of SO$_2$. Relative to previous Sonora models, we also update the line list sources for NH$_3$ and CH$_4$. For completeness, all opacity sources are summarized in Table \ref{tab:opatab}. Our opacity tables are computed on a 1460 pressure–temperature grid that covers temperatures 75–4000~K and pressures 1$\times 10^{-6}$ - 3000 bars. For each atmospheric layer, \texttt{PICASO} estimates the opacity at that layer’s pressure and temperature by identifying the surrounding grid points in pressure-temperature space and interpolating between them. The interpolation is performed in logarithmic pressure and inverse temperature, $(1/T)$, rather than directly in pressure and temperature. In the limited cases where a layer lies outside the tabulated opacity grid, \texttt{PICASO} uses the nearest two grid points at the edge of the table and extends the same interpolation formula beyond the grid boundary. Further details on the different opacity treatments can be found in \citet{Mang2026}.

For FeH, we follow the treatment adopted in previous Sonora models by reducing the opacity in the H band ($\sim 1.6 \upmu$m) by a factor of three to better reproduce observed spectra. While this empirical correction significantly improves agreement with observations, recent work by \citet{Mader2026} has identified persistent wavelength-dependent discrepancies, including outside the H band, when comparing \texttt{Sonora Diamondback} models to spectra of late-M and L dwarfs. This suggests that improved FeH line lists will be required in future work. 

The neutral alkali opacities are treated separately from the molecular line lists because their pressure-broadened resonance lines can strongly shape the red optical and near-infrared spectra of brown dwarfs. We use the same neutral alkali metal atomic line absorption calculations for Li, Na, K, Rb, and Cs as in \texttt{Sonora Bobcat} \citep{Marley2021}. For the Na and K line opacities, the updated opacity set used incorporates newer unified alkali line profiles from \citet{Allard2016, Allard2019}. These profiles capture the far wings of the Na and K resonance doublets in H$_2$/He-rich atmospheres, where standard Lorentzian or molecular pressure-broadening prescriptions are insufficient. The effects of pressure-broadened alkali opacity are most important across the $Y$ and $J$ bands.

The H$_2$ collision-induced absorption (CIA) opacity also depends on the assumed ortho-to-para ratio of H$_2$. Normal H$_2$ corresponds to a fixed 3:1 ortho-to-para ratio, while equilibrium H$_2$ follows the local Boltzmann population and becomes increasingly para-rich at low temperatures. As described in \citet{Freedman2008}, the H$_2$ CIA opacity used here assumes normal H$_2$. The choice of ortho-to-para ratio, which can be affected by disequilibrium processes (conceptually reminiscent of CO to $\rm CH_4$ conversion), is most important in cold atmospheres, such as Uranus \citep{Smith1978orthopara, MassieHuntsen1982H2orthapara, Fegley1991uranus, MarleyMcKay1999}, and thus affects the coldest models in our grid. More work is needed to explore the impact of equilibrium and normal ortho--para H$_2$ populations on the H$_2$ CIA opacity and resulting spectra of our models.

We also distinguish between the treatment of opacities in the equilibrium and disequilibrium grids. For the equilibrium grid, we use pre-mixed (pre-weighted) correlated-$k$ tables defined on the standard 1460-point P–T grid and a 196-wavelength grid, constructed using precomputed equilibrium chemistry abundances. The disequilibrium grid uses individual molecular correlated-$k$ tables on the same 1460-point P–T grid, but over a 661-wavelength grid. In this case, opacities are combined “on-the-fly” for each model using the resort–rebin method of \citet{Amundsen2017}, rather than using on pre-mixed chemistry tables. Further details on these methods and associated opacity updates in \texttt{PICASO} are provided in \citet{Mang2026}.

Using these opacities together with the 1D thermal structures, we compute moderate-resolution thermal emission spectra with \texttt{PICASO}. The spectra span 0.5–30 $\upmu$m at a resampled resolving power of $R = 60{,}000$, which is sufficiently high to be readily binned and convolved to be used even at the highest resolution of JWST observations. The publicly available spectra on Zenodo are provided at a resolution of $R = 30{,}000$ to reduce storage requirements and improve accessibility.

\subsubsection{Chemistry}\label{sec:chemistry}

The chemical equilibrium abundances are based on thermochemical models from \citet{Lodders2002, Lodders2006} and \citet{Visscher2006, Visscher2010, Visscher2012}. Our metallicity grid spans [M/H] = -1.0 to +2.0, where +0.0 is solar metallicity. For each metallicity, we also have four different C/O ratios of 0.229, 0.458, 0.687, and 1.14, where solar is 0.458 \citep{lodders2009}. While the solar C/O ratio has been updated in \citet{lodders2020}, we continue to use 0.458 as the solar value to be consistent with previous Sonora grids of models.

For disequilibrium models, the strength of vertical mixing in the atmosphere is parameterized by the eddy diffusion coefficient ($K_{\rm zz}$). While the parameterization of $K_{\rm zz}$ remains largely uncertain, the impact of different prescriptions for $K_{\rm zz}$ has been evaluated in \citet{Mukherjee2022}. For consistency with \texttt{Sonora Elf Owl}, we adopt the same constant $K_{\rm zz}$ values of $10^2, 10^4, 10^7, 10^8, \rm and~10^9$. These values set the characteristic mixing timescale $\tau_{\rm mix}$ in each atmospheric layer,

\begin{equation}
    \tau_{\rm mix} = \dfrac{H^2}{K_{\rm zz}}
\end{equation}

where $H$ is the local pressure scale height. Larger values of $K_{\rm zz}$ correspond to more efficient mixing and therefore shorter mixing timescales and a lower value of $K_{\rm zz}$ results in a slower mixing timescale.

\begin{figure}
    \centering
    \includegraphics[width=\columnwidth]{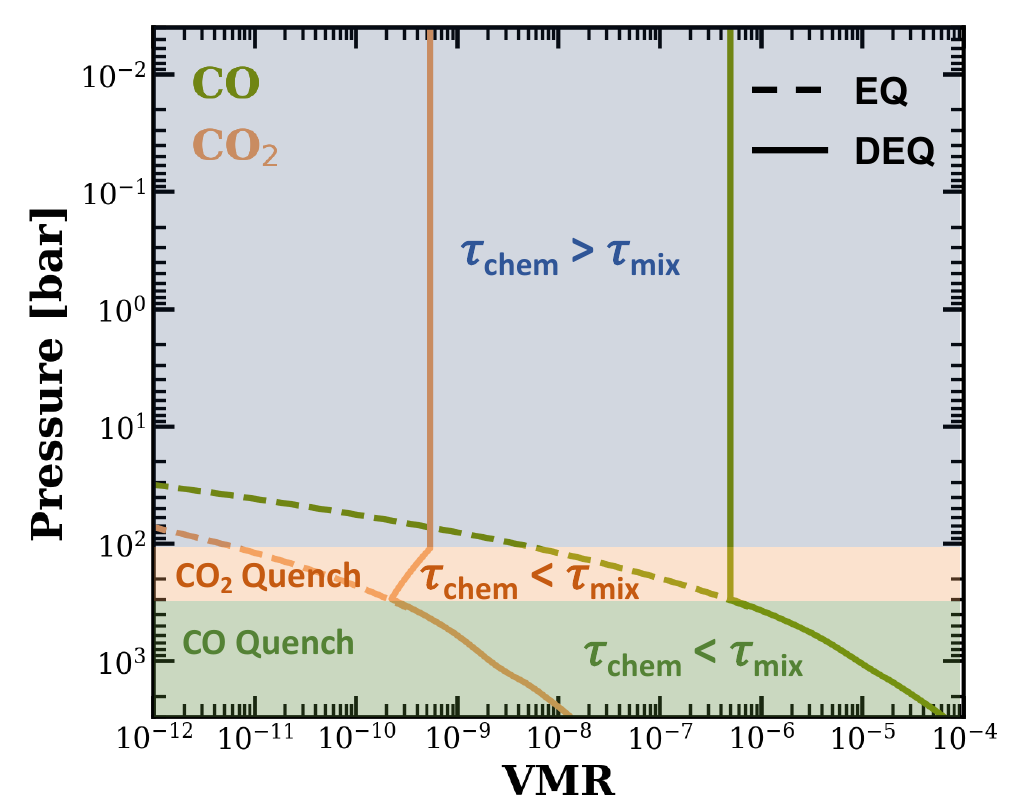}
    \caption{Diagram showing the different quench points of CO and CO$_2$ in a 275 K brown dwarf with log(g) = 4.5, [M/H] = +0.0, C/O = 0.458, and $K_{\rm zz} = 10^7$. CO (green line) quenches first deeper in the atmosphere. Then CO$_2$ (orange line) has a second quench further up in the atmosphere (orange region) as it continues to react and equilibrate with the quenched CO abundances. Finally, both abundances are held constant in the regions where mixing dominates (blue region).}
    \label{fig:quench}
\end{figure}

To determine where species such as CH$_4$, CO, CO$_2$, H$_2$O, NH$_3$, N$_2$, and HCN depart from chemical equilibrium, the mixing timescale in each layer is compared to the chemical timescale, $\tau_{\rm chem}$, associated with the relevant reactions (Figure~\ref{fig:quench}). In the deep atmosphere, where temperatures and pressures are high, chemical reactions proceed rapidly ($\tau_{\rm chem} \ll \tau_{\rm mix}$), and abundances follow equilibrium values as seen in the green region of Figure \ref{fig:quench}. Moving up in the atmosphere towards cooler temperatures and lower pressures, the reaction rates decrease and a transition occurs when $\tau_{\rm chem} \gtrsim \tau_{\rm mix}$. Above this level, vertical mixing outpaces chemical conversion, and the abundances become “quenched.” We define the quench point at this transition and hold the abundances constant above it. This is the prescription followed in both \texttt{Sonora Cholla} and \texttt{Sonora Elf Owl}. 

\begin{figure}
    \centering
    \includegraphics[width=\columnwidth]{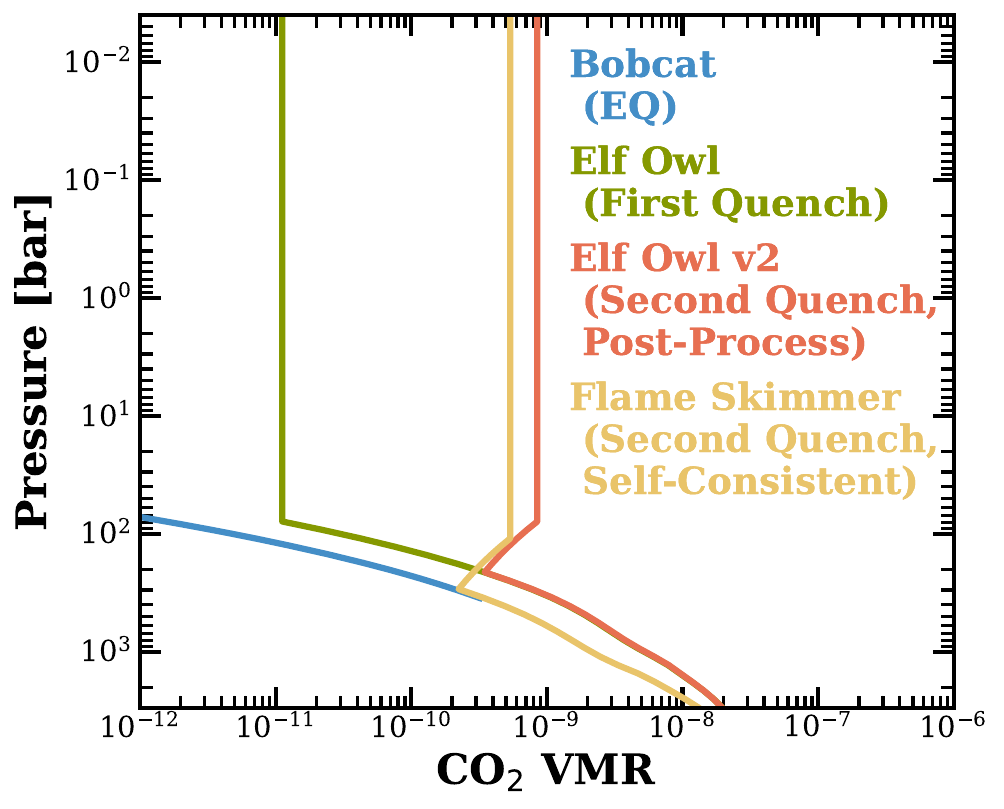}
    \caption{Volume mixing ratios of CO$_2$ in a 275 K object with log(g) = 4.5, [M/H] = +0.0, C/O = 0.458, and for disequilibrium models, $K_{\rm zz} = 10^9$. The profiles for \texttt{Sonora Bobcat}, \texttt{Sonora Elf Owl}, \texttt{Sonora Elf Owl v2}, and \texttt{Sonora Flame Skimmer} are shown in different colors. The updated CO$_2$ kinetics in \texttt{Sonora Flame Skimmer} produce significantly higher CO$_2$ abundances due to the second quench and it is important to calculate these self-consistently.}
    \label{fig:co2}
\end{figure}

Our disequilibrium models introduce three key refinements relative to \texttt{Cholla} and \texttt{Sonora Elf Owl}. First, we revise the quenching prescription for CO$_2$, CH$_4$, H$_2$O, CO, and HCN. For CO$_2$, previous \texttt{Sonora Elf Owl} models have been shown to underestimate its abundance relative to JWST observations of substellar atmospheres \citep{Beiler2024}. Revisiting the chemical kinetic framework of \citet{Visscher2010Icarus} and \citet{Zahnle2014} revealed that the CO$_2$ abundance profile naturally exhibits a “second quench point” (Figure~\ref{fig:quench}, orange region). This arises because CO quenches deeper in the atmosphere (Figure~\ref{fig:quench}, green region), setting an enhanced reservoir of CO that can subsequently be converted into CO$_2$ before CO$_2$ itself quenches at lower pressures. As a result, the CO$_2$ abundance is increased by one to two orders of magnitude relative to models that do not account for this effect (Figure~\ref{fig:co2}). Although \texttt{Sonora Elf Owl v2} \citep{Wogan2025} implemented this CO$_2$ correction, the thermal structure was not recomputed, resulting in a non-self-consistent atmospheric profile (Figure \ref{fig:co2}, red line). The difference between the CO$_2$ abundance in \texttt{Sonora Flame Skimmer} (yellow) and \texttt{Sonora Elf Owl v2} (red) arises from modest changes to the P–T structure.

For other species like CH$_4$, H$_2$O, CO, and HCN, these chemical reaction timescales were assumed to be independent of metallicity in \texttt{Sonora Elf Owl}. While the effects of this assumption are minimal, for completeness, in the \texttt{Sonora Flame Skimmer} disequilibrium models, the chemical timescale used to calculate the quench point for these species follows the metallicity-dependent relationships prescribed in \citet{Zahnle2014}. For CO, CH$_4$, and H$_2$O, the chemical timescale follows eq. 12 in \citet{Zahnle2014} :

\begin{equation}
    \tau_{\rm chem CO/CH_4} = 1.5 \times 10^{-6} P^{-1} m^{-0.7} \rm exp(42,000/T)
\end{equation}

where $P$ is the pressure in bars, $m$ is the linear metallicity relative to solar (ie. [M/H] = +0.5 is $m \sim$ 3), and $T$ is the temperature in Kelvin. For HCN, the chemical timescale follows eq. 40 in \citet{Zahnle2014}:

\begin{equation}
    \tau_{\rm HCN} = 1.5 \times 10^{-4} P^{-1} m^{-0.7} \rm exp(36,000/T).
\end{equation}

\begin{figure*}
    \centering
    \includegraphics[width=\linewidth]{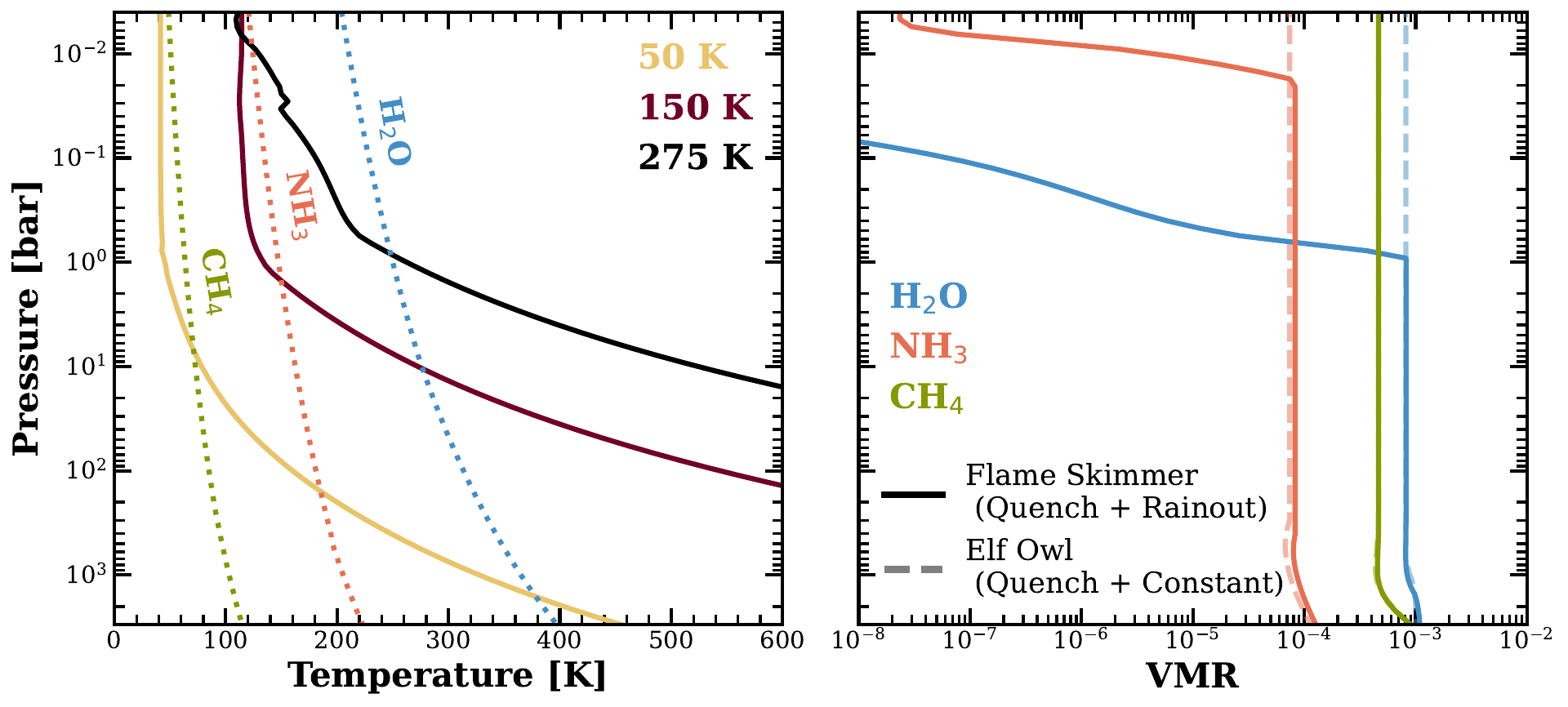}
    \caption{Left: Solid lines show the pressure-temperature profiles of three objects at 50, 150, and 275~K with log(g) = 4.5, [M/H] = +0.0, C/O = 0.458, and $K_{\rm zz} = 10^7$. The dashed lines correspond to the condensation curves of CH$_4$, NH$_3$, and H$_2$O from left to right. Right: volume mixing ratios of H$_2$O, NH$_3$, and CH$_4$ in the 275 K object. Solid lines show the condensation of volatiles in \texttt{Sonora Flame Skimmer}, compared to the constant quenched abundances used in \texttt{Sonora Elf Owl} (dashed). The depletion of these volatile species above their condensation levels is important to accurately model compositions of cold substellar objects.}
    \label{fig:chemistry}
\end{figure*}

Second, we allow volatile species to condense and be depleted from the gas phase, following the rainout chemistry approach used in \texttt{Sonora Bobcat}. All equilibrium chemistry models already include this treatment, whereas the disequilibrium prescription in \texttt{Sonora Elf Owl} held quenched abundances fixed above the quench level (Figure~\ref{fig:chemistry}, right). As a result, species such as H$_2$O, NH$_3$, and CH$_4$ remained overabundant in the coldest atmospheres, where they are expected to condense out of the gas phase.

Because \texttt{Sonora Flame Skimmer} extends to temperatures as low as 50~K, we allow H$_2$O, NH$_3$, and CH$_4$ to condense after quenching by limiting their gas-phase abundances to the corresponding saturation vapor pressures (Figure~\ref{fig:chemistry}, left). This treatment modifies the atmospheric composition by removing excess volatile gas, thereby avoiding unrealistically strong absorption features that would otherwise be predicted for these cold atmospheres. This treatment modifies the atmospheric composition but does not include cloud opacity. For disequilibrium models colder than 450~K, these species are also ``cold-trapped,'' such that once condensation begins, their abundances are not permitted to increase relative to the layer below.

Finally, we remove PH$_3$ from the disequilibrium models. The phosphorus chemical network and reaction kinetics remain uncertain in cold substellar atmospheres \citep{Wang2016, visscher2020, Bains2023ESC.....7.1219B}. The PH$_3$ abundances predicted by \texttt{Sonora Elf Owl} were found to be inconsistent with observations of cool brown dwarfs \citep{Beiler2024}. Based on this observational evidence, \texttt{Sonora Elf Owl v2} \citep{Wogan2025} removed PH$_3$ as a spectral contributor. Recent observations have reported PH$_3$ detections in WISE~0855--0714 at the $\sim$1~ppb level \citep{Rowland2024} and in Wolf~1130C at $\sim$0.100~ppm \citep{Burgasser2025}. For comparison, theoretical disequilibrium chemistry calculations predict PH$_3$ abundances approximately 5$\times$ higher than inferred for Wolf~1130C and 500$\times$ higher than inferred for WISE~0855--0714. Recent calculations show the depletion of phosphorus could be through condensation with Fe and Ni to form metal phosphides \citep{Yin2026}. For uniformity across the grid and following the prescription used in \texttt{Sonora Elf Owl v2}, we therefore exclude PH$_3$ from the default chemistry. Since PH$_3$ is a trace species, its omission is not expected to significantly affect the thermal structure of the models, and PH$_3$ opacity can be incorporated in post-processing for custom models in future work.

\subsection{Evolution Models}\label{sec:evolution}

The evolution models used are described in detail in \citet{chachan2025} and we summarize the key features and differences here. Our evolution models solve the standard structure equations using the relaxation method. The atmospheric models presented in this work are used as boundary conditions at the `top' of the planet's interior, i.e., the relation of the surface gravity and interior entropy to outgoing flux that sets the cooling of an object. Following \citet{chachan2025}, we define the atmosphere--interior boundary at 10~bar. At each timestep, the interior model is matched to the atmospheric grid by requiring the specific entropy of the convective interior to agree with the entropy of the atmospheric model at this pressure level. In this way, the atmosphere and interior are stitched together through a common entropy at 10~bar, while the emergent flux from the atmospheric model determines the rate at which the interior cools. The radii reported in the evolutionary tracks correspond to the atmospheric radius at 10~mbar. 

The interior consists of a homogeneous and adiabatic envelope whose composition is set by the metallicity of the atmospheric model used for the boundary condition. Objects $< 20~M_\oplus$ are assumed to have a $12~M_\oplus$ core, objects $\geq 20~M_\oplus$ but $< 3~M_{\rm J}$ have a $15~M_\oplus$ core, and beyond $3~M_{\rm J}$ the objects are assumed to be coreless. The choice of core masses is motivated by recent empirical constraints on giant-planet mass--radius relationships. In particular, \citet{chachan2025} infer a characteristic core mass of $M_{\rm core}=14.7^{+1.8}_{-1.6}~M_\oplus$. We therefore adopt a 15~$M_\oplus$ core for most low-mass giant-planet tracks. For the lowest-mass objects in the grid, however, we adopt a slightly smaller 12~$M_\oplus$ core. This choice avoids an unrealistically core-dominated structure for $< 20~M_\oplus$ models while preserving the expected trend that lower-mass planets have larger bulk heavy-element fractions. The cooling of the core contributes to the luminosity. 

Metals are represented by water and the EOS from \citet{Mazevet2019} is used. For hydrogen and helium, we use the EOS from \citet{chabrier2021}, which incorporates improvements in our understanding of their behavior at intermediate pressure levels ($0.1 - 10$ Mbar) and accounts for the effect of non-ideal mixing of H and He. The water and H-He EOS are combined assuming they constitute an ideal mixture. These choices depart from the Bobcat evolution models, which used the SCvH EOS for H and He and represented metals by excess He. This change in EOS is also accompanied by small changes in X = 0.725 and Y = 0.275 to match the tables available for the latest EOS. Solar metallicity is assumed to be Z = 0.014, in line with our atmospheric models. The D/H mass ratio = $2.88 \times 10^{-5}$ is kept identical to previous models. However, all mass ratios are calculated accounting for the assumed Z. 

The atmospheric model grid presented in this work extends up to $T_{\rm eff}=2400$~K. However, at early ages and for the most massive objects, the evolutionary tracks require high-entropy initial conditions corresponding to effective temperatures above this upper limit. We do not compute additional atmospheric models above 2400~K because this regime requires a stellar-atmosphere treatment with additional atomic opacities not included in our atmospheric model framework. Instead, for these high-temperature initial conditions, we extend the atmospheric boundary condition above 2400~K in order to initialize the cooling tracks. \cite{Davis2025} show that linear extrapolation of models curves can significantly overestimate $T_{\rm eff}$ at early ages. We choose to fit a simple parametric model to SPHINX models used in \cite{Davis2025} to extend our grid for $T_{\rm eff} > 2400$ (see Appendix~\ref{sec:SPHINX_fits}). This parametric model includes a dependence on log~$g$ and atmospheric metallicity, which is used to extend the smaller range of these parameters in the SPHINX models to match our atmospheric models.

We model the evolution of objects spanning a broad range of masses from 15~$M_\oplus$ to $0.1~M_\odot$. In particular, we extend the mass grid of the Bobcat models to include both lower mass objects (15, 20, 40, 60 $M_\oplus$ and 0.25 and 0.75 $M_{\rm J}$) that may be observed in upcoming observing campaigns as well as objects more massive than $0.08~M_\odot$ to enable modeling of objects that straddle the hydrogen burning limit. The evolution models are run up to an age of 10 Gyr and are only terminated earlier if the lowest entropy in the EOS is reached (this only affects the lowest mass objects).

\section{Results} \label{sec:results}

\subsection{P-T Profiles} \label{sec:ptprofiles}

\begin{figure*}
    \centering
    \includegraphics[width=\linewidth]{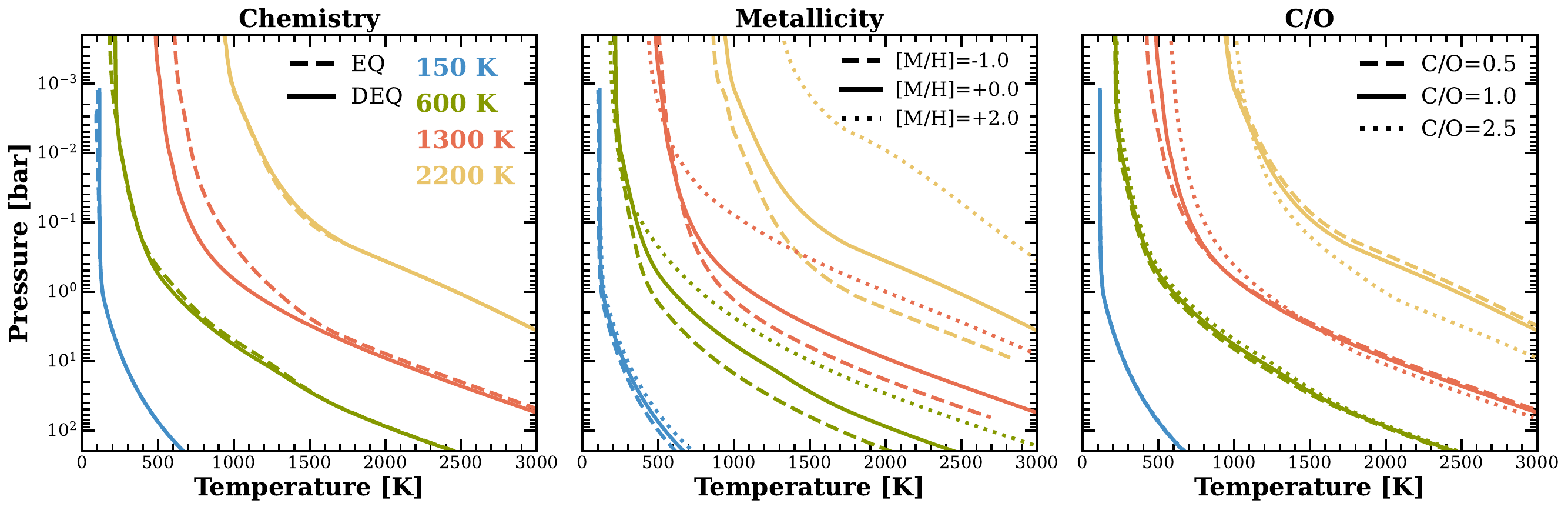}
    \caption{Pressure-temperature profiles of a sequence of \texttt{Sonora Flame Skimmer} objects with $T_{\rm eff} = 150, 600, 1300,\rm{and}~2200$~K, log(g) = 4.5, and $K_{\rm zz} = 10^7$. Left: comparison of equilibrium (dashed) and disequilibrium (solid) chemistry models, showing only minor differences, with a shift in the radiative–convective boundary near 1300~K. Middle: models with varying metallicity, [M/H] = $-1.0$ (dashed), $+0.0$ (solid), and $+2.0$ (dotted), where increasing metallicity shifts the entire P–T profile to warmer temperatures. Right: models with different C/O ratios, 0.229 (dashed), 0.458 (solid), and 1.14$\times$ solar (dotted), where higher C/O ratios produce a colder convective region. Overall, metallicity has the dominant impact on the thermal structure, while disequilibrium chemistry and C/O introduce comparatively smaller modifications.}
    
    \label{fig:pt_sequence}
\end{figure*}

Figure \ref{fig:pt_sequence} shows a range of P-T profiles in the \texttt{Sonora Flame Skimmer} grid and the effect that varying the parameters of the grid has at different temperatures. The profiles shown are for 150, 600, 1300, and 2200~K objects with log(g) = 4.5. The solid lines are the same profiles across all the panels. In the first panel on the left, we compare equilibrium models (dashed) with disequilibrium models with $K_{\rm zz} = 10^7$ (solid). The largest difference is seen in the 1300~K model. The difference arises from the equilibrium model having a deeper convective zone compared to the disequilibrium model. This temperature lies near the L/T transition and near the CO/CH$_4$ chemical transition, where vertical mixing will significantly modify the gas-phase abundances of these dominant carbon-bearing species. These abundance changes alter the wavelength-dependent opacity and will shift the location of the radiative--convective boundary, producing a larger change in the thermal structure than is seen at the other temperatures shown. At higher temperatures, carbon remains primarily in CO, and the equilibrium and disequilibrium abundance structures are more similar. 

In the middle panel, we show models with different metallicities. We show the extreme cases of the grid with models with metallicities of [M/H] = -1.0 (dashed) being shifted to be colder compared to the solar metallicity models (solid). This monotonic trend continues as increasing the metallicity, up to 100$\times$ solar (dotted), shows the shift of the P-T profiles to much hotter temperatures due to the higher opacities. Finally, in the right panel we show the effects of different C/O values. Variations in C/O have a comparatively small effect on the P--T profiles. Unlike metallicity, which changes the total abundance of heavy elements and therefore broadly increases molecular opacity, changing C/O at fixed metallicity primarily redistributes carbon and oxygen among major opacity sources such as H$_2$O, CO, CH$_4$, and CO$_2$. As a result, the thermal profiles remain relatively similar across the C/O range explored here. In the extreme cases of 0.5 times solar (dashed) and 2.5 times solar (dotted), the major difference comes at higher temperatures, with higher C/O leading to cooler deep atmospheres but hotter upper atmospheres, with the opposite effect for lower C/O ratios.

We also compared the parts of the grid that overlap with the \texttt{Sonora Bobcat} and \texttt{Sonora Elf Owl} grid to validate our fiducial cases, given the changes we have made. \texttt{Sonora Flame Skimmer} reproduces the \texttt{Sonora Bobcat} P–T structures with high fidelity. This agreement validates our use of \texttt{PICASO} to generate the equilibrium grid, demonstrating consistency with the original \texttt{Sonora Bobcat} models computed using the Fortran \texttt{EGP} code as we expand this grid to more extreme metallicities (10x sub-solar and 100x super-solar). We also find minimal differences with the \texttt{Sonora Elf Owl} grid except at cold temperatures, where differences in the upper atmosphere arise due to the inclusion of volatile condensation in \texttt{Sonora Flame Skimmer}. For further details and discussion, see Appendix \ref{sec:bobcat_elf_pt}.

\subsection{Spectral Features} \label{sec:spectra}

\begin{figure*}
    \centering
    \includegraphics[width=\textwidth]{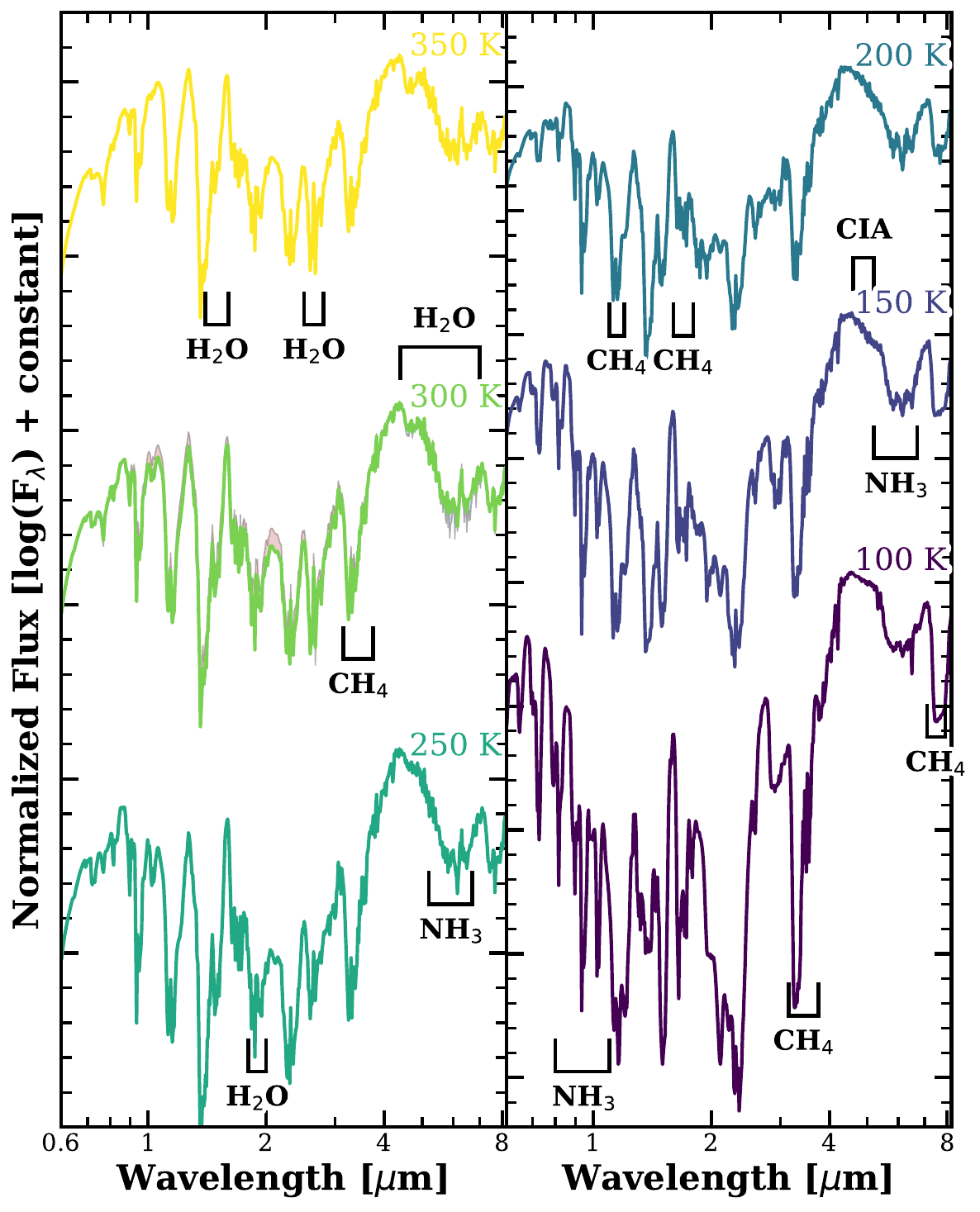}
    \caption{Cold spectral sequence of \texttt{Sonora Flame Skimmer} models for objects with $T_{\rm eff} = 350$-$100$~K, $\log(g) = 4.5$, [M/H] = +0.0, C/O = 0.458, and $K_{\rm zz} = 10^7$. \texttt{Sonora Elf Owl} spectra (black) are shown for comparison at 300 and 350~K. Shaded regions highlight differences between the models, where blue indicates stronger absorption (lower flux) in \texttt{Sonora Elf Owl} relative to \texttt{Sonora Flame Skimmer}, and red indicates weaker absorption (higher flux). These differences arise from the inclusion of volatile condensation and updated disequilibrium chemistry in \texttt{Sonora Flame Skimmer}, which reduces upper-atmosphere opacities and alters spectral features at the coldest temperatures.}
    \label{fig:coldspec}
\end{figure*}

Figure~\ref{fig:coldspec} shows a sequence of emergent spectra from 350~K down to 100~K at fixed gravity (log($g$)=4.5), solar metallicity, and $K_{\rm zz}=10^7$. The corresponding \texttt{Sonora Elf Owl} spectra at 300 and 350~K (black) are also plotted for a direct comparison at these overlapping temperatures. At $T_{\rm eff} = 300$~K, there is a clear effect from the hybrid disequilibrium chemical treatment of the volatiles. The \texttt{Sonora Elf Owl} spectrum is fainter, particularly in the H$_2$O band (5-7$\upmu$m), due to stronger absorption from H$_2$O, which remains at a constant abundance from the quench level. In contrast, by allowing volatiles to condense in \texttt{Sonora Flame Skimmer}, it depletes these volatile species in the upper atmosphere, reducing opacity and increasing the emergent flux in these absorption windows. As the sequence progresses to colder temperatures ($T_{\rm eff} \lesssim 200$~K), the impact of volatile condensation becomes more pronounced. These trends highlight the importance of condensing volatiles when modeling ultra-cool atmospheres, even when considering disequilibrium chemistry.

\begin{figure*}
    \centering
    \includegraphics[width=\textwidth]{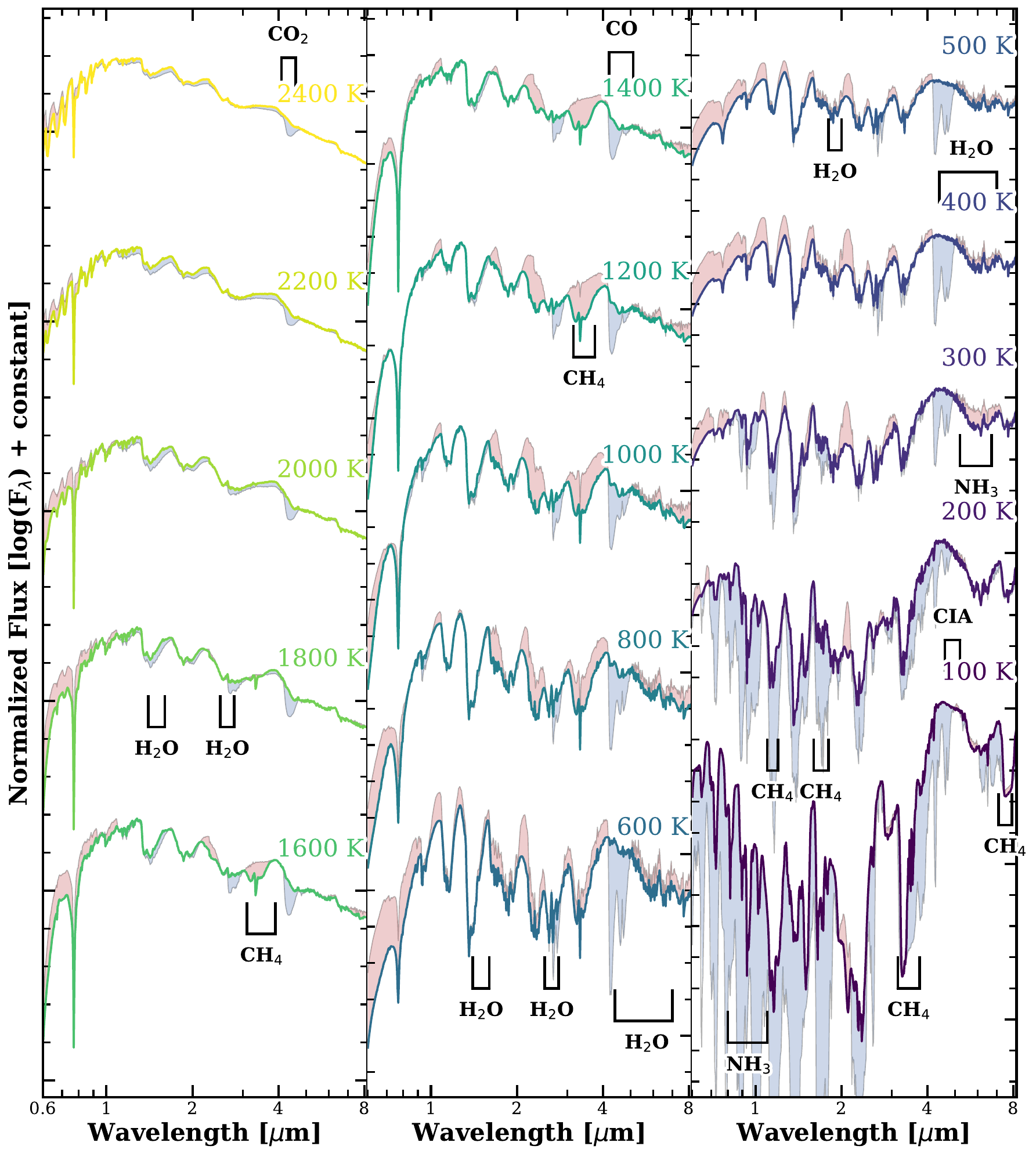}
    \caption{Spectral sequence for \texttt{Sonora Flame Skimmer} models spanning $T_{\rm eff} = 2400$-$100$~K with $\log(g) = 4.5$, [M/H] = +0.0, C/O = 0.458, and $K_{\rm zz} = 10^7$. These are compared to metal-rich models with [M/H] = +2.0 (black). Shaded regions highlight differences between the spectra, where blue indicates stronger absorption (lower flux) and red indicates weaker absorption (higher flux) in the high-metallicity models relative to solar metallicity. Increasing metallicity enhances molecular opacities across the atmosphere, leading to systematically suppressed flux and stronger absorption features.}
    \label{fig:mhspec}
\end{figure*}

The other key comparison is the differences in metallicity in the models. Figure~\ref{fig:mhspec} isolates the effect of metallicity by comparing spectra at [M/H] = $+0.0$ (colored lines) and [M/H] = $+2.0$ (black lines). Increasing metallicity produces stronger molecular absorption features, most notably from CO$_2$ in the 4--4.5~$\upmu$m region. In addition to CO$_2$, metallicity also affects the balance between CH$_4$ and CO. At intermediate temperatures ($T_{\rm eff}$ \textless~1600~K), higher metallicity models exhibit reduced CH$_4$ absorption relative to solar metallicity models. This behavior reflects the warmer P-T profiles at high metallicity, which shift the chemistry toward CO-dominated compositions prior to quenching. At the coldest temperatures ($T_{\rm eff} \lesssim 200$~K), these differences diminish as the thermal structures converge. In this regime, spectral differences are dominated by variations in H$_2$O abundance, which increases with metallicity and strengthens absorption features across the near- and mid-IR.

\subsection{Evolutionary Tracks} \label{sec:evoresults}

\begin{figure*}
    \centering
    \includegraphics[width=\linewidth]{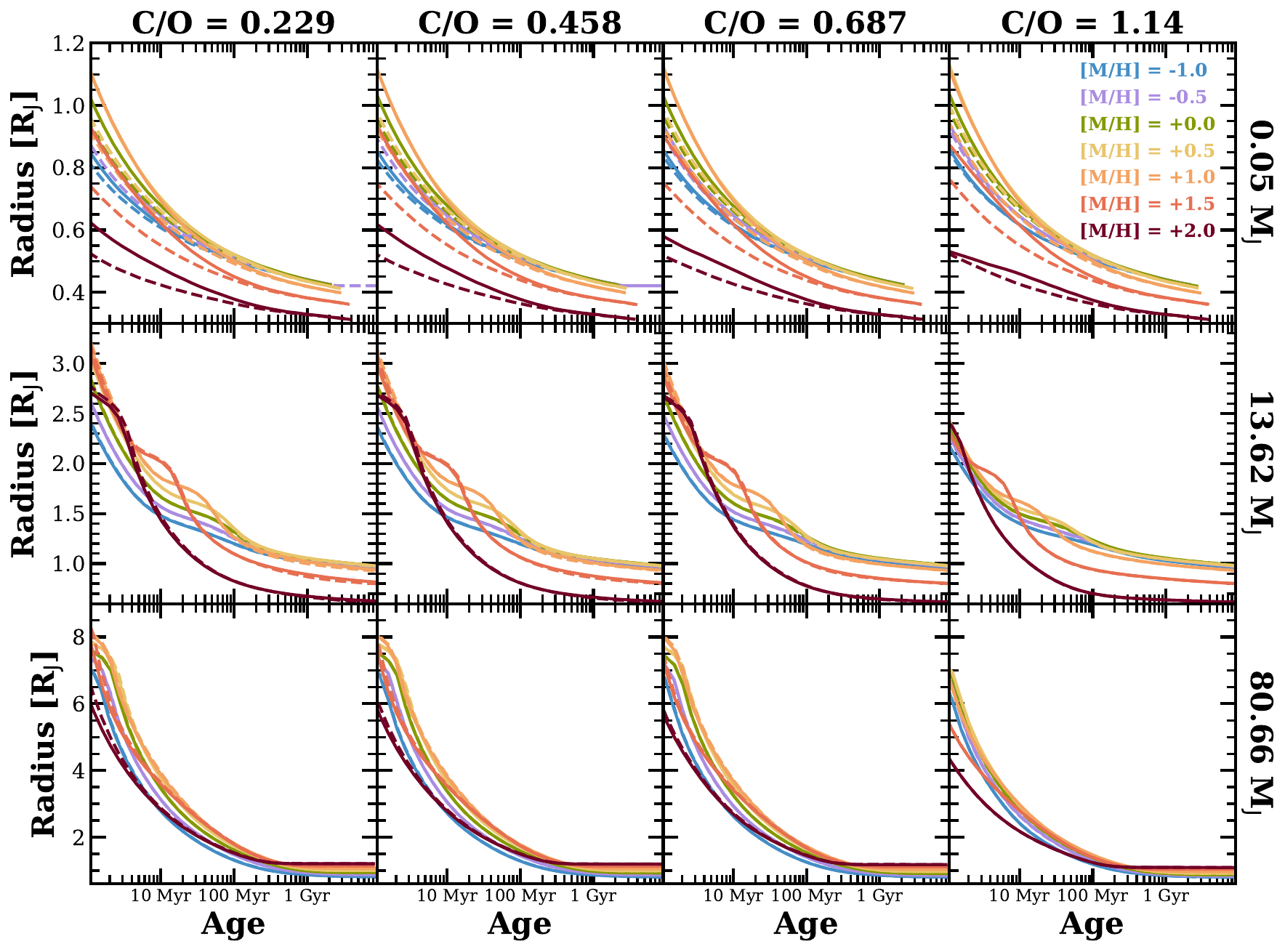}
    \caption{Radius as a function of time for different C/O ratios (columns) and three representative substellar masses, low mass (0.05 $M_{\rm J}$), near the deuterium-burning limit (13.62 $M_{\rm J}$), and high mass (80.66 $M_{\rm J}$) (rows). Colors indicate atmospheric metallicity. Solid lines show equilibrium chemistry models, while dashed lines correspond to disequilibrium models with $K_{\rm zz} = 10^7$. Metallicity drives the dominant variations in radius evolution, influencing both cooling rates at early times and the onset of nuclear burning, while variations in C/O only have a minor impact. Disequilibrium chemistry has a larger effect at low masses.}
    
    \label{fig:evolmodelspanel}
\end{figure*}

Figure~\ref{fig:evolmodelspanel} shows the evolution of radius as a function of time for three different masses for the range of C/O and metallicity of our atmospheric models. Atmospheric models with equilibrium chemistry are shown in solid lines, and those with disequilibrium chemistry with an assumed $K_{\rm zz} = 10^7$ are shown in dashed lines. Evolutionary models using disequilibrium chemistry atmospheric boundary conditions show little variation with the choice of $K_{\rm zz}$, and we therefore present results for a single representative value of $K_{\rm zz}$. 

Differences between the equilibrium and disequilibrium tracks are most pronounced at early times in the highest-metallicity cases, while becoming progressively smaller at lower metallicities, though measurable differences in the radius remain. They also appear only at the lowest masses, where the cooling history is most sensitive to the atmospheric boundary condition. In these cold, low-luminosity objects, changes in opacity from disequilibrium chemistry generate a thermal profile shifted colder than the equilibrium chemistry model and produce smaller radii at young ages. At higher masses, the larger interior entropy reservoir and nuclear energy release dominate over these atmospheric chemistry effects, leading to nearly identical equilibrium and disequilibrium tracks.

More metal-rich objects are expected to be smaller due to their higher density at any given pressure and temperature, and this is mostly borne out by the evolution models with two interesting exceptions. Firstly, at early ages, intermediate metallicity objects are larger than the lowest metallicity objects in our model grid. This is due to delayed cooling of the metal-rich objects as a result of additional opacity from metals in the atmosphere. However, the effect of increasing metallicity on the opacity typically does not dominate over the effect on density for atmospheric metallicity of $+2.0$. Secondly, at mature ages, objects for which hydrogen fusion becomes important (i.e. $\gtrsim 80~M_{\rm J}$) show the opposite trend with more metal-rich objects remaining larger than metal-poor objects on many Gyrs. This is a result of an earlier onset of hydrogen burning in more metal-rich objects, which stalls the cooling and contraction of the envelope. This is also evident in Figure~\ref{fig:evolmodelspanel}, which shows the upturn in radius for the most massive sub-stellar objects occurs at earlier times for higher atmospheric metallicities.

\begin{figure}
    \centering
    \includegraphics[width=\columnwidth]{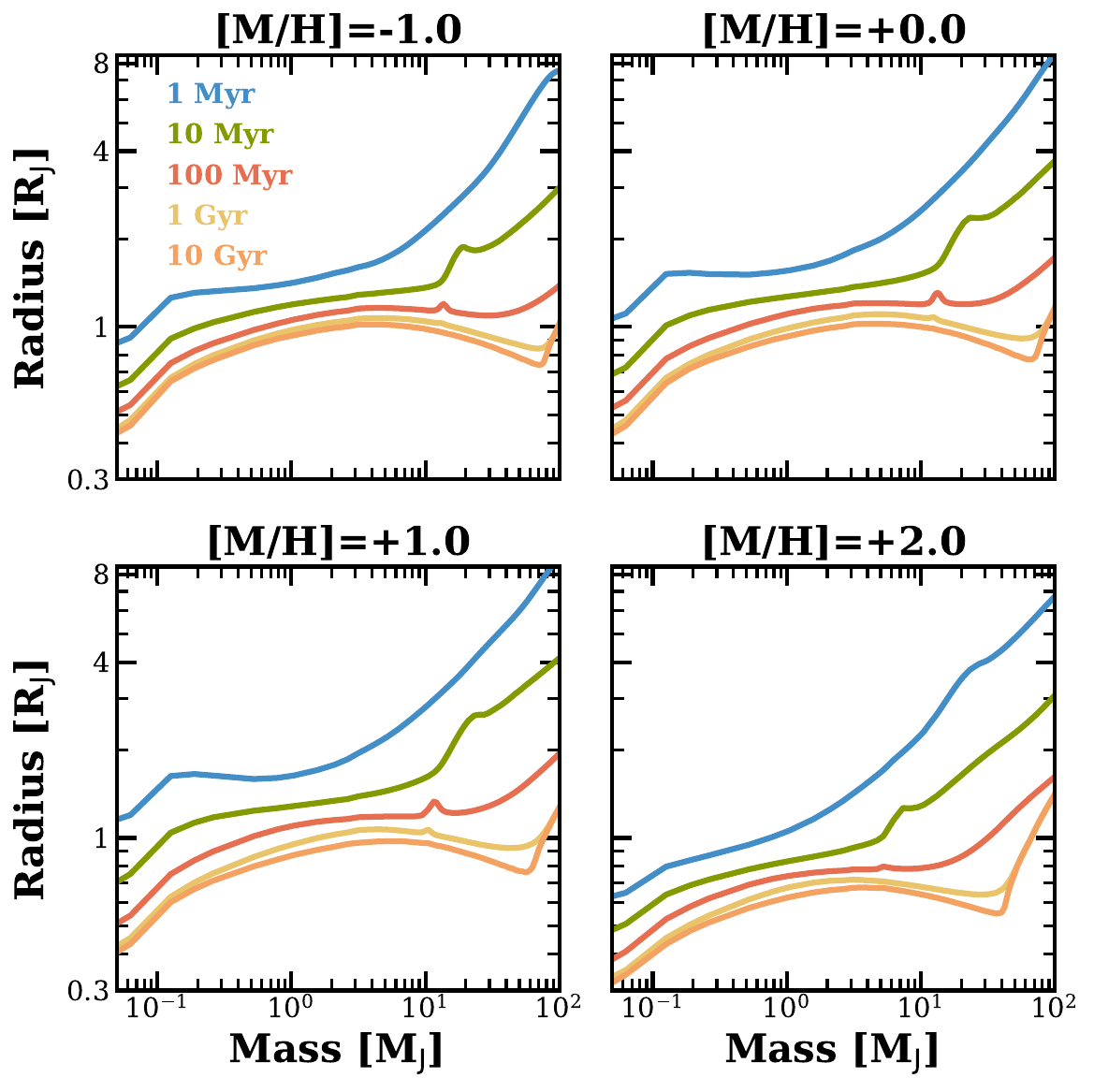}
    \caption{Mass–radius relation of substellar objects at different ages for four atmospheric metallicities, with C/O = 0.458. Higher metallicity shifts the onset of deuterium burning to earlier times and lower masses, producing larger radii at young ages and extending deuterium burning to objects as low as $\sim 5~M_{\rm J}$.}
    \label{fig:evolutionisochrones}
\end{figure}

Metallicity also has an impact on the onset of deuterium burning, both the timing and the object mass at which the onset happens. In the middle row of Figure~\ref{fig:evolmodelspanel} corresponding to a $13.62~M_{\rm J}$ mass object, deuterium-burning (manifest in the slowing down of contraction and radius evolution) turns on earlier for more metal-rich objects. Figure~\ref{fig:evolutionisochrones} shows the mass-radius relation of substellar objects for different metallicities at different times in their evolutionary sequence. The panel corresponding to a metallicity of +2.0 shows that deuterium-burning starts earlier (the 1 Myr curve shows signs of deuterium burning) and also extends to lower mass objects down to $\sim 5~M_{\rm J}$. 

Table \ref{tab:dbmm} shows the deuterium-burning minimum mass (DBMM) and hydrogen-burning minimum mass (HBMM) for the equilibrium \texttt{Sonora Flame Skimmer} evolution models as a function of metallicity. We report two DBMM values, defined as the masses at which objects have burned 50\% and 90\% of their initial deuterium by an age of 10~Gyr. The HBMM is defined as the mass at which 99.9\% of the luminosity is powered by nuclear fusion. For both DBMM and HBMM, increasing metallicity systematically lowers the minimum mass. 

\begin{table}[]
    \centering
    \begin{adjustbox}{width=1.18\columnwidth, right}
    \begin{tabular}{|c|c|c|c|}
    \hline
    \textbf{{[M/H]}} &  \textbf{DBMM (50\%)} & \textbf{DBMM (90\%)} & \textbf{HBMM} \\
    \hline
    -1.0 & 13.01 & 13.86 & 81.18 \\
    \hline
    -0.5 & 12.66 & 13.46 & 79.59 \\
    \hline
    +0.0 & 12.21 & 12.98 & 77.92 \\
    \hline
    +0.5 & 11.44 & 12.20 & 75.01 \\
    \hline
    +1.0 & 10.05 & 11.08 & 68.65 \\
    \hline
    +1.5 & 7.85  & 8.39  & 57.45 \\
    \hline
    +2.0 & 5.39  & 6.11  & 45.03 \\
    \hline
    \end{tabular}
    \end{adjustbox}
    \caption{Deuterium Burning Minimum Mass and Hydrogen Burning Minimum Mass in Jupiter masses from the equilibrium evolution tracks as a function of metallicity with solar C/O.}
    \label{tab:dbmm}
\end{table}

At solar metallicity, our DBMM is 0.05~$M_{\rm J}$ higher than that derived from the clear \texttt{Sonora Diamondback} models, while the sub-solar case ([M/H] = $-0.5$) is only 0.02~$M_{\rm J}$ lower. At higher metallicity (3$\times$ solar), the DBMM increases by 0.12~$M_{\rm J}$ relative to \texttt{Sonora Diamondback}. For the 90\% depletion criterion, we recover the same DBMM (12.9~$M_{\rm J}$) as reported for \texttt{Sonora Diamondback} and \texttt{Sonora Bobcat}. In comparison to \citet{Speigel2011}, our DBMM values most closely align with the He28 models, although they are lower by $\sim$0.6~$M_{\rm J}$. The He28 models adopt a helium mass fraction of $Y=0.28$, which is close to the value used here ($Y=0.275$) and to the solar value of $Y=0.274$ \citep{Lodders2003}. The remaining differences likely arise from updates to the atmospheric opacities and the equation of state (EOS). 

For the HBMM at solar metallicity, our value (77.92~$M_{\rm J}$) is higher than the \texttt{Sonora Diamondback} (76.5~$M_{\rm J}$) and \texttt{Sonora Bobcat} (77.5~$M_{\rm J}$) results, and lies slightly below the values reported by \citet{Saumon2008} and \citet{Chabrier2023} (78.6~$M_{\rm J}$). This suggests that the updated EOS from \citet{chabrier2021} shifts the hydrogen-burning limit to higher masses. At the highest metallicity, the HBMM is at a strikingly low mass of only 45.03~$M_{\rm J}$, almost half the value of the HBMM at 10x sub-solar metallicity. The effect of metallicity found here is even stronger than the impact of irradiation on transiting brown dwarfs, which can reduce the HBMM and DBMM by 13\% and 16\%, respectively, depending on the strength of irradiation \citep{Mukherjee2026}.

\begin{figure*}
    \centering
    \includegraphics[width=0.48\linewidth]{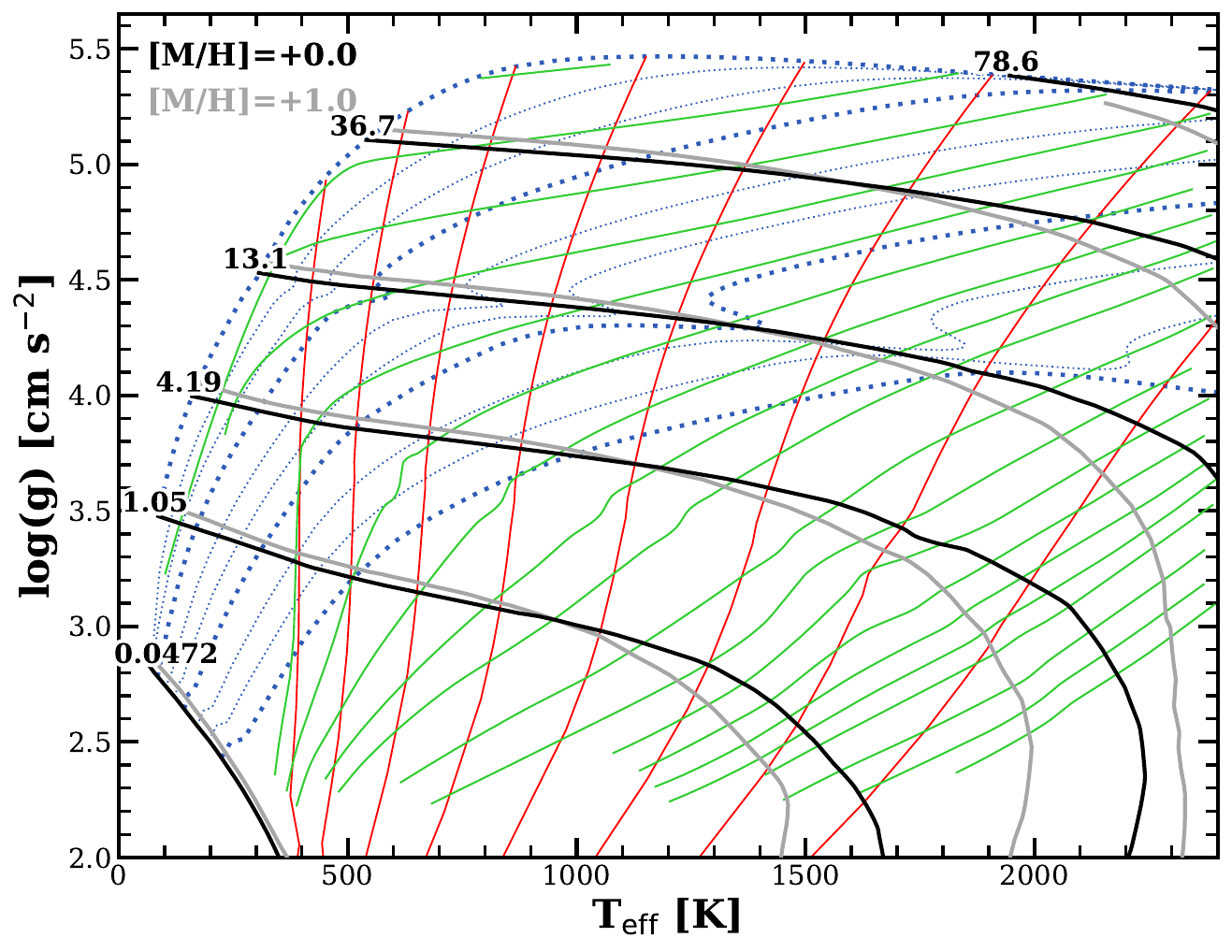}
    \includegraphics[width=0.48\linewidth]{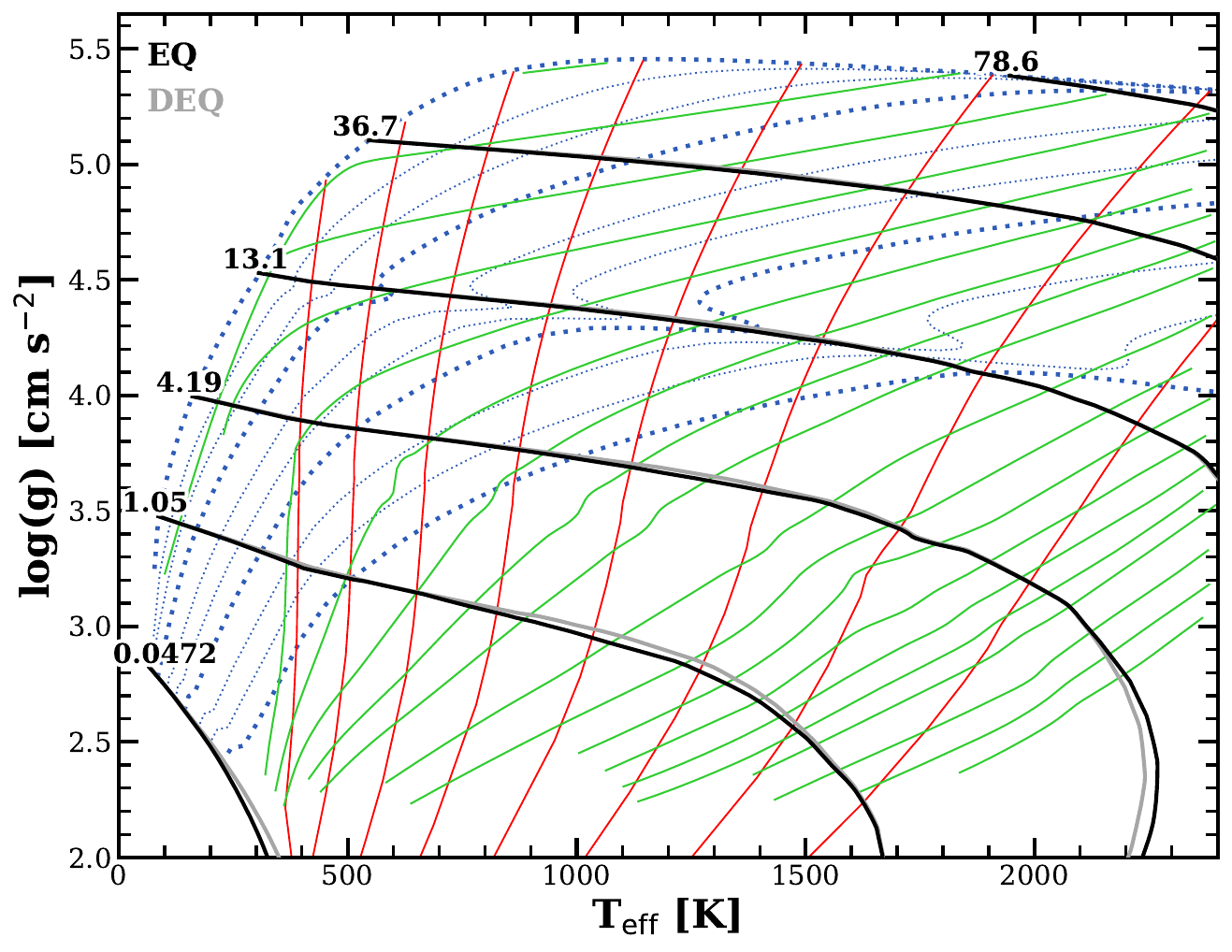}
    \caption{Evolution of brown dwarfs in $T_{\rm eff}$ and surface gravity for different masses. Evolution proceeds from right to left along the thick black tracks, labeled by mass in Jupiter masses. Isochrones (blue dotted lines) span 10~Myr to 10~Gyr, with thicker lines marking 0.01, 0.1, 1, and 10~Gyr. Nearly vertical red lines indicate constant luminosity from $\log(L/L_\odot) = -3$ to $-6.5$ in steps of 0.5, and green lines show constant radii from 0.08 to 0.42~$R_\odot$. Left: solar metallicity tracks (black) compared to enhanced metallicity ([M/H] = +1.0; gray). Right: equilibrium chemistry models (black) compared to disequilibrium models (gray). Metallicity produces noticeable shifts in the evolutionary tracks, particularly at higher temperatures, while differences between equilibrium and disequilibrium chemistry remain small.}
    \label{fig:sm08}
\end{figure*}

Figure~\ref{fig:sm08} shows the evolution of substellar objects in $T_{\rm eff}$–$\log(g)$ space, where individual mass tracks progress from high temperatures and low surface gravities toward lower temperatures and higher gravities as the objects contract and cool over time. The kink visible along the isochrones at intermediate masses marks the onset of deuterium burning, which temporarily slows contraction and produces a plateau in both $T_{\rm eff}$ and $\log(g)$. The $13.1~M_{\rm J}$ track, near the deuterium-burning limit, clearly traces this feature across temperature space at solar metallicity. We also note that some regions of the grid are included to maintain a uniform parameter space but do not correspond to physically plausible objects within the age of the universe (e.g., $T_{\rm eff} = 100$~K at $\log(g)=5.5$) where no mass track reaches this parameter space. These edge cases are retained for completeness but should be interpreted with caution; further discussion is provided in Section~\ref{sec:convergence}.

The left panel of Figure \ref{fig:sm08} highlights the impact of metallicity on the evolutionary tracks. At fixed mass and age, metal-rich models are systematically shifted to lower $T_{\rm eff}$ and slightly lower $\log(g)$ at early times, reflecting slower cooling due to enhanced atmospheric opacity. This delay in cooling is most pronounced at higher temperatures. Additionally, the onset of deuterium burning occurs earlier and extends to lower masses for higher metallicity objects, consistent with the increased opacity and modified internal structure. As the objects age, these differences diminish as cooling dominates over the initial atmospheric conditions, and a higher metallicity track, at a fixed mass and age, leads to slightly warmer and higher surface gravity objects.

In contrast, the right panel demonstrates that differences between equilibrium and disequilibrium chemistry have only a minor effect on the bulk evolutionary tracks. While disequilibrium chemistry can significantly alter molecular abundances and the observed spectra, its impact on the cooling history is comparatively small. As a result, the equilibrium and disequilibrium tracks largely overlap across most of parameter space, with only subtle deviations appearing at lower masses and higher temperatures. This behavior underscores that metallicity and opacity dominate the thermal evolution, while disequilibrium chemistry plays a secondary role, primarily affecting observable spectral features rather than the global cooling sequence. Further comparisons with \citet{Saumon2008} evolutionary tracks can be found in Appendix \ref{sec:sm08_appendix}.

\begin{figure}
    \centering
    \includegraphics[width=\columnwidth]{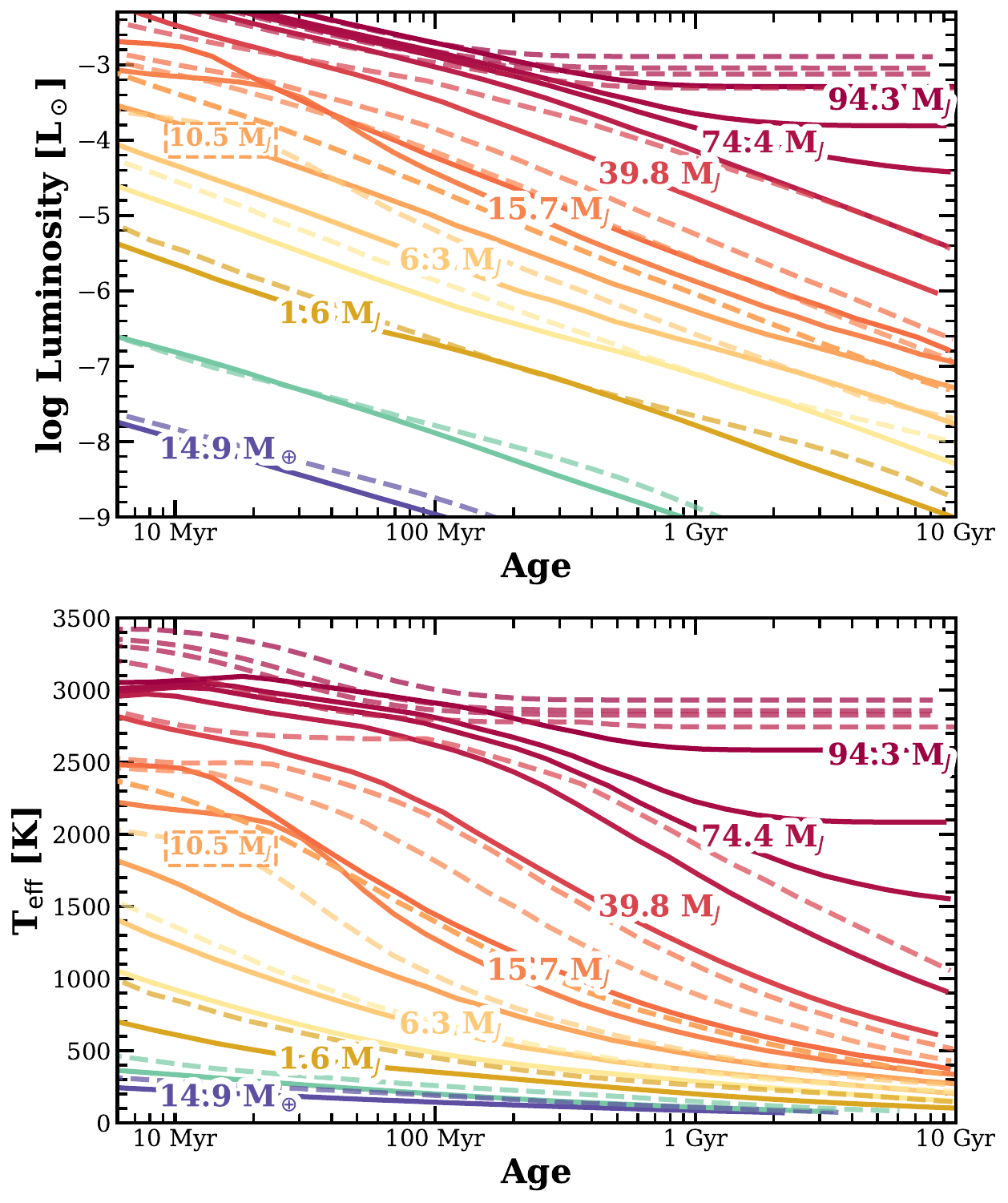}
    \caption{Thermal evolution models for the equilibrium \texttt{Sonora Flame Skimmer} grid comparing solar ([M/H] = +0.0; solid) and super-solar ([M/H] = +2.0; dashed) metallicity tracks in luminosity (top) and effective temperature (bottom) as a function of age. Higher metallicity slows the cooling of substellar objects, leading to higher luminosities and effective temperatures at fixed age, with the largest effects at higher masses where nuclear burning processes are modified.}
    \label{fig:evolution}
\end{figure}

Looking further into the effects of metallicity, Figure~\ref{fig:evolution} shows the evolution of substellar objects in both luminosity and effective temperature as a function of age. Increasing metallicity slows the cooling of these objects, resulting in higher luminosities and effective temperatures at a given age. For intermediate mass objects, metallicity also shifts the onset of deuterium burning to earlier times. For example, the evolutionary track of a 15.7~$M_{\rm J}$ object at solar metallicity intersects that of a 10.5~$M_{\rm J}$ object at [M/H] = +2.0 around $\sim$20~Myr. By this time, the lower-metallicity object has cooled significantly, with a temperature difference of $\sim$600~K relative to its higher-metallicity counterpart. 

At even higher masses, enhanced metallicity increases the temperatures in the deep atmosphere of the object, increasing the efficiency of hydrogen burning via the pp-chain, allowing objects to sustain fusion at lower masses (down to $\sim$55.5~$M_{\rm J}$ in this grid). Higher metallicity atmospheres also retain heat more efficiently, further extending the duration over which fusion can be sustained. While the extreme case of 100$\times$ solar metallicity represents an upper bound, it illustrates the broader trend that metallicity strongly influences the thermal and nuclear evolution of massive substellar objects. 

\begin{figure}
    \centering
    \includegraphics[width=\columnwidth]{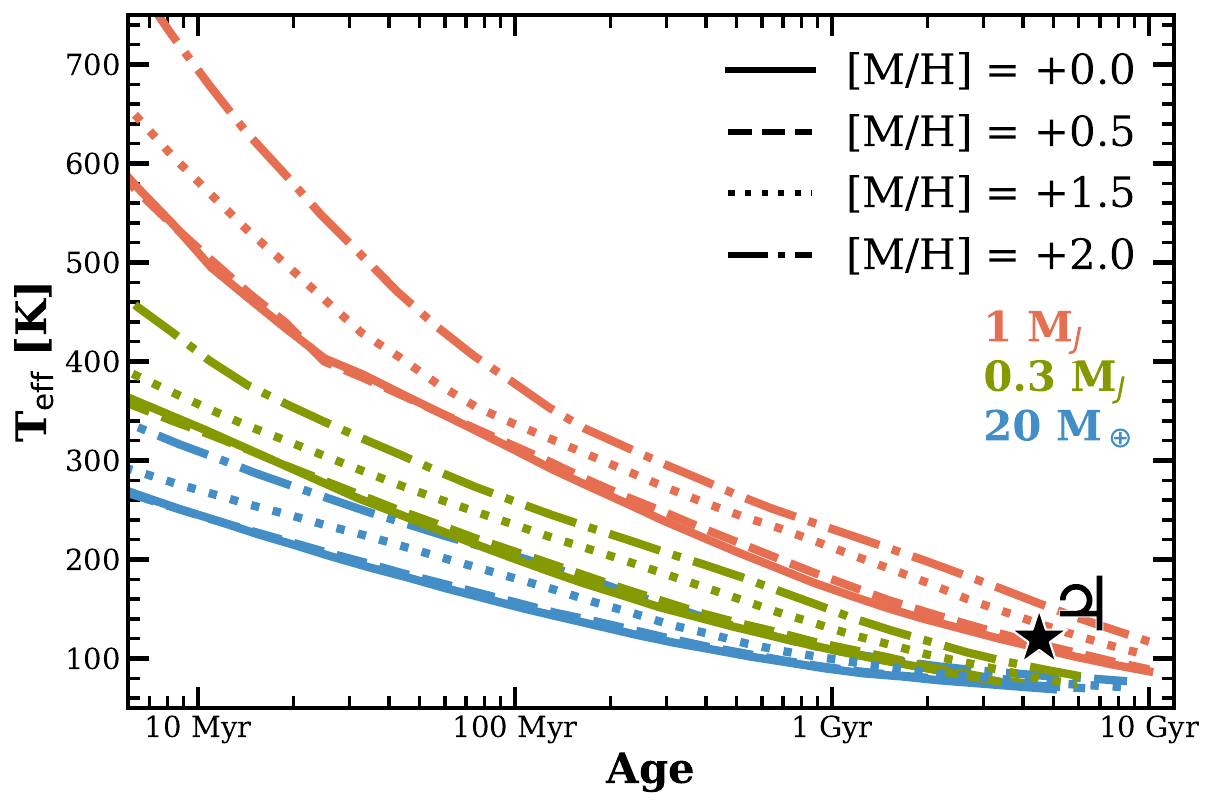}
    \caption{Thermal evolution models for the equilibrium \texttt{Sonora Flame Skimmer} grid for objects with masses similar to Neptune (blue), Saturn (green), and Jupiter (red). Different metallicities are shown from solar (solid) to 100$\times$ super-solar (dot-dashed). Jupiter is shown as a black star for reference. Higher metallicity slows the cooling of these objects, producing temperature differences of $\gtrsim$150~K at young ages and $\sim$50~K at late times for Jupiter-mass objects, with smaller differences at lower masses.}
    
    \label{fig:evolutionlowmass}
\end{figure}

For low-mass objects, Figure~\ref{fig:evolutionlowmass} shows cooling tracks at varying metallicities for objects with masses comparable to Neptune, Saturn, and Jupiter. At all three masses, differences between solar and moderately enhanced metallicities (e.g., 3$\times$ solar) are small, but these differences become increasingly pronounced at higher metallicities and younger ages. For a Neptune-mass object, the maximum temperature difference at 10~Myr is $\sim 60$~K, decreasing to negligible differences at late times. This effect grows with mass: a Jupiter-mass object exhibits differences of up to $\sim 200$~K at 10~Myr, which diminish to $\sim 40$~K at older ages.

These trends reflect the increased atmospheric opacity at higher metallicities, which slows radiative cooling and delays contraction, particularly at early times when objects are most luminous. As the objects age, the influence of atmospheric composition weakens, and the tracks converge. Overall, metallicity introduces a mass-dependent modulation of the early thermal evolution, with the largest impact for higher-mass objects and at young ages, with the cooling tracks converging as a function of time.

\subsubsection{Comparison with Bobcat models and effect of modifications}

\begin{figure*}
    \centering
    \includegraphics[width=0.48\linewidth]{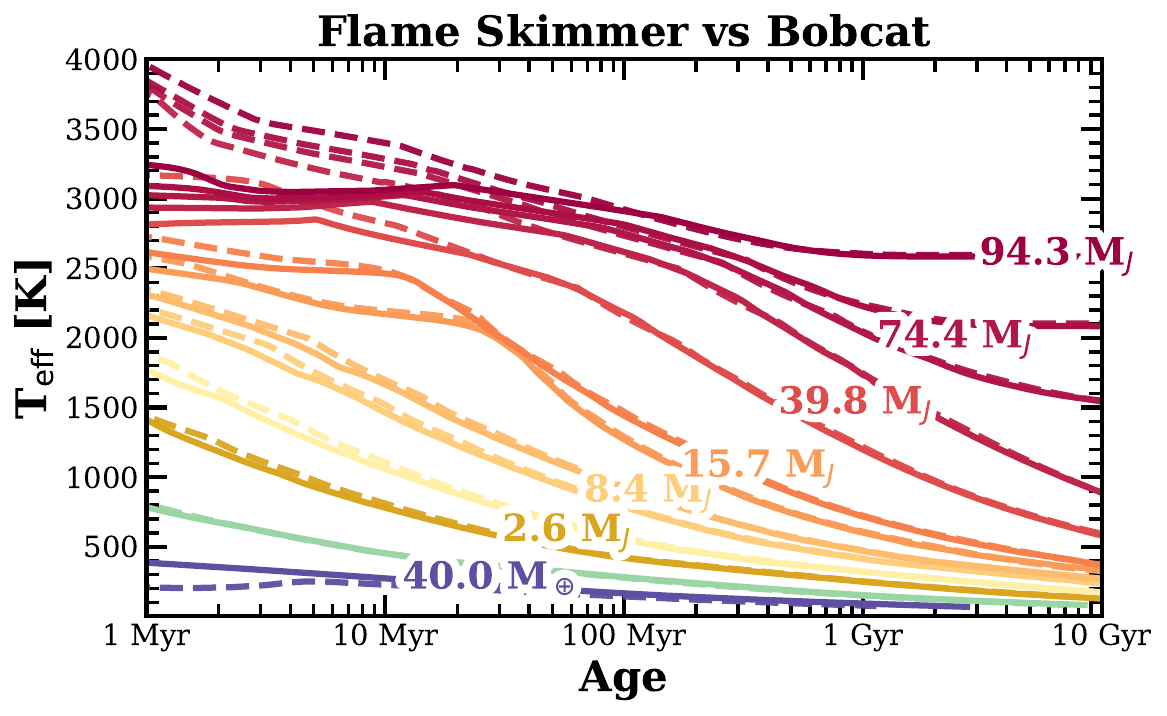}
    \includegraphics[width=0.48\linewidth]{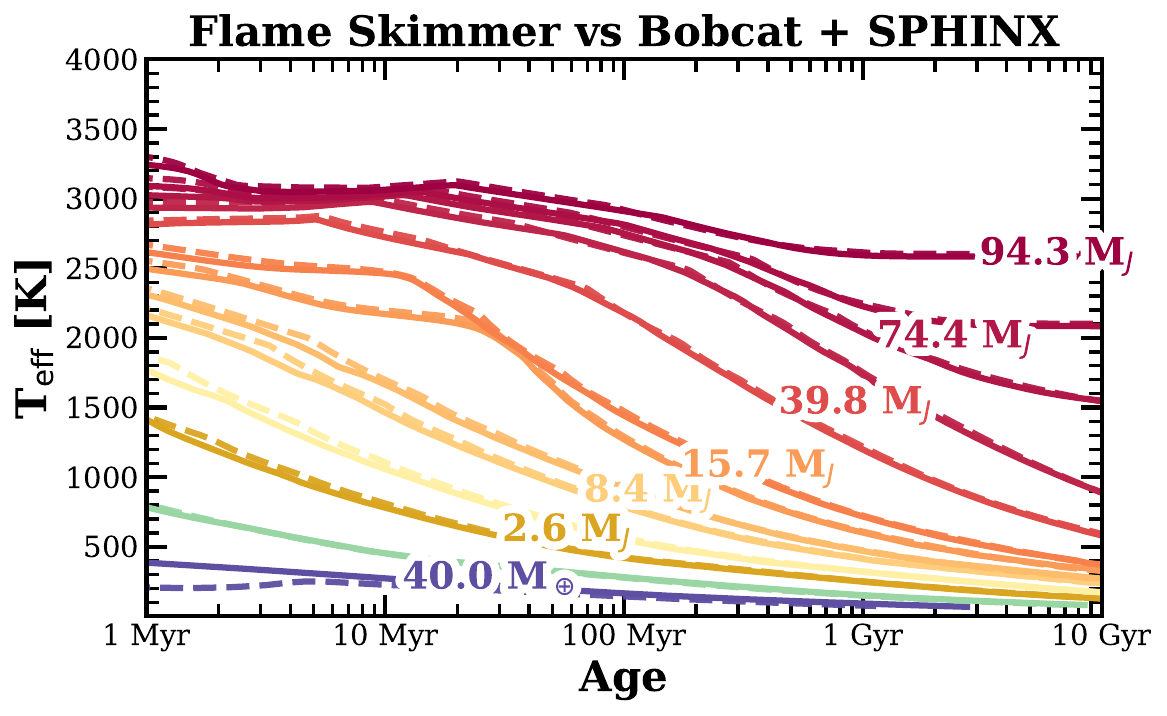}
    \caption{Left: comparison of \texttt{Sonora Bobcat} (dashed) and \texttt{Sonora Flame Skimmer} (solid) evolutionary models for solar metallicity and solar C/O objects. Right: effect of adopting SPHINX models for the atmospheric boundary condition in the \texttt{Sonora Bobcat} models (dashed). The \texttt{Sonora Flame Skimmer} models closely reproduce the original \texttt{Sonora Bobcat} evolution, with the largest differences arising from the choice of atmospheric boundary condition at higher temperatures for young.}
    \label{fig:bobcat_comparison}
\end{figure*}

We compare the evolutionary models presented here with the \texttt{Sonora Bobcat} evolution models for a solar metallicity and solar C/O atmosphere. To do so, we use our evolution code to replicate the \texttt{Sonora Bobcat} evolution model setup. Our reproduction of \texttt{Sonora Bobcat} models agrees well with the models presented in \cite{Marley2021} with small differences for objects at the deuterium-burning limit around 10 Myr. These differences become insignificant at later ages. We study the effect of each update that we made to the evolution models in this paper by adding this change to the Bobcat models one at a time. Figure~\ref{fig:bobcat_comparison} shows comparisons of evolution models for a representative set of planet masses for the most notable differences.

The left panel in Figure \ref{fig:bobcat_comparison} shows a direct comparison of the resulting evolutionary tracks with effective temperature as a function of age between the choices made in \texttt{Sonora Flame Skimmer} (solid lines) and \texttt{Sonora Bobcat} (dashed lines). Overall, the \texttt{Sonora Flame Skimmer} and \texttt{Sonora Bobcat} tracks do not have large differences at mature ages. At the high substellar mass range, the difference between the two evolutionary tracks is driven by how the atmospheric boundary condition is extended beyond its available range (Figure \ref{fig:bobcat_comparison}; right panel). \texttt{Sonora Bobcat} models assumed linear extrapolation for $T_{\rm eff}$ and $T_{10}$, which severely overestimates $T_{\rm eff}$ as $T_{10}$ increases and leads to rapid cooling at early ages and thus smaller radii compared to the models in this paper \citep{Davis2025}. We choose to use the SPHINX models to inform the extrapolation of our atmospheric boundary conditions. The objects in our models more closely follow the Hayashi track, and their $T_{\rm eff}$ remains nearly constant while they cool and become smaller. This choice in using the SPHINX models reconciles the largest differences between the \texttt{Sonora Flame Skimmer} and \texttt{Sonora Bobcat} models at younger ages.

\begin{figure}
    \centering
    \includegraphics[width=\columnwidth]{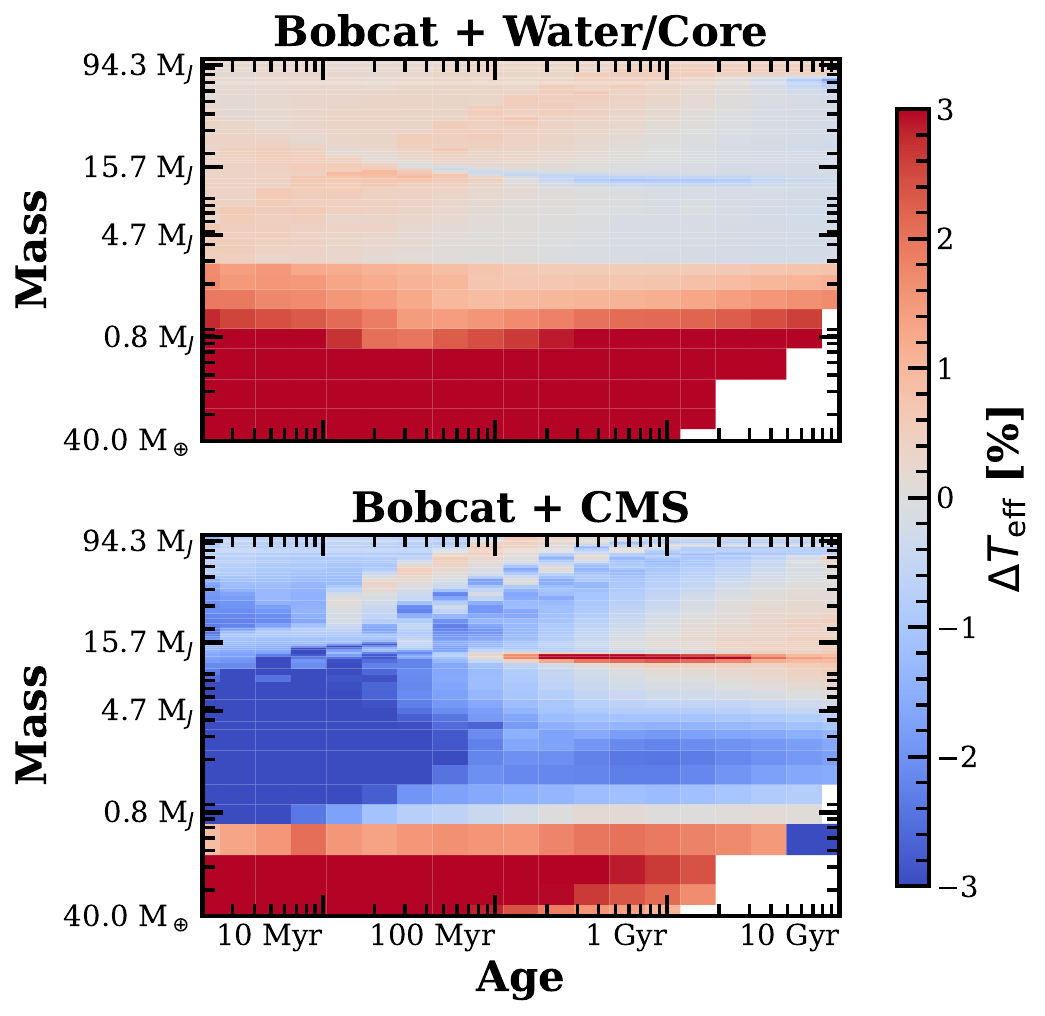}
    \caption{Diagnostic heat maps showing the fractional difference in effective temperature ($\Delta T_{\rm eff}$) between modified \texttt{Sonora Bobcat} evolutionary models and the original \texttt{Sonora Bobcat} tracks as a function of mass and age. Each panel isolates the impact of a single model update: Top: inclusion of a water-rich core in low-mass objects; Bottom: replacement of the SCvH equation of state with the CMS equation of state for H–He. Colors indicate the percent change in $T_{\rm eff}$ relative to the original \texttt{Sonora Bobcat} models. These updates produce modest ($\lesssim$ few percent) shifts in $T_{\rm eff}$.}

    \label{fig:bobcatchanges}
\end{figure}

Other updates to the \texttt{Sonora Flame Skimmer} evolutionary models introduce comparatively minor differences relative to the inclusion of SPHINX atmospheric boundary conditions at higher temperatures. Figure~\ref{fig:bobcatchanges} shows the change in effective temperature, $\Delta T_{\rm eff}$, between the original \texttt{Sonora Bobcat} evolutionary models and models with individual updates applied. Negative values (blue) indicate models that are cooler at a given mass and age, while positive values (red) indicate models that are warmer relative to the original tracks. The horizontal feature near $\sim 12$-$15~M_{\rm J}$ traces the deuterium burning limit, which introduces a localized change in the thermal evolution. The magnitude of these temperature differences is small compared to the deviations introduced by the SPHINX boundary condition at young ages. At the lowest masses, the dominant effect arises from the inclusion of a $12$-$15~M_\oplus$ core (Figure~\ref{fig:bobcatchanges}, top panel), which increases the bulk metallicity of sub-Jovian objects and leads to smaller radii. Replacing the SCvH equation of state with the CMS equation of state for H--He (Figure~\ref{fig:bobcatchanges}, bottom panel) produces a similar shift toward smaller radii, with the largest impact in the sub-Saturn to Saturn mass range. In contrast, changes associated with using water as a proxy for metals, as well as adjustments to the H, He, and metal mass fractions, produce only minor differences and are therefore not shown. These effects are particularly small at solar metallicity and tend to partially cancel. The impact of modifying the atmospheric boundary condition (excluding SPHINX) is also omitted, as it introduces only subtle changes to the evolutionary tracks.

\subsection{Color-Magnitude Diagram} \label{sec:cmd}

\begin{figure*}
    \centering
    \includegraphics[width=\textwidth]{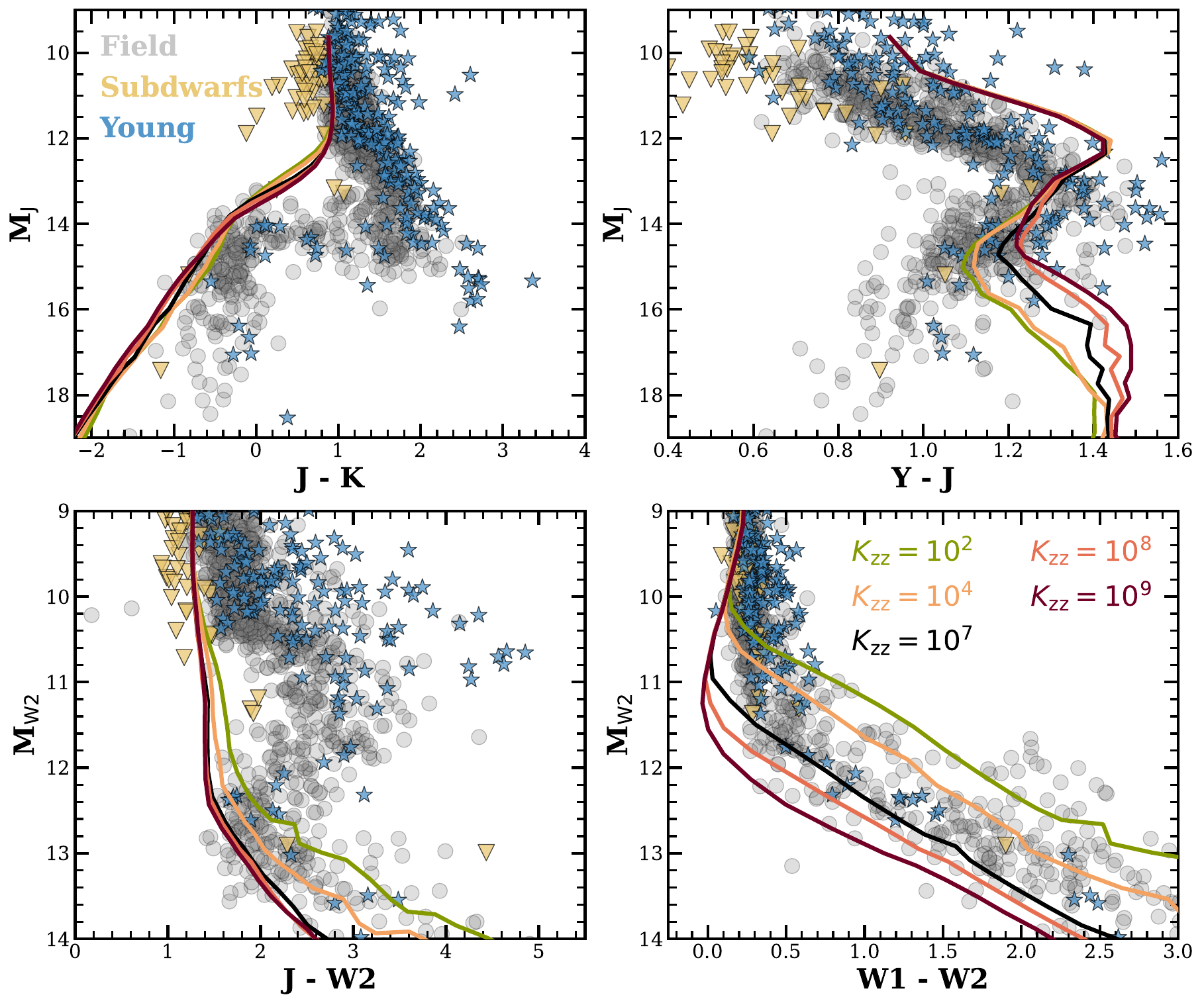}
    \caption{Color–magnitude diagrams using MKO and WISE (W1, W2) filters, with a comparison sample drawn from the UltracoolSheet. \texttt{Sonora Flame Skimmer} disequilibrium tracks are shown for $\log(g)=4.5$, [M/H] = +0.0, and C/O = 0.458, with varying $K_{\rm zz}$. Variations in vertical mixing produce only minor shifts in these MKO filters, indicating that $K_{\rm zz}$ has a limited impact on broadband photometry at these wavelengths compared to in W1 and W2.}
    \label{fig:cmdkzz}
\end{figure*}

Figure~\ref{fig:cmdkzz} shows color--magnitude diagrams (CMDs) for the \texttt{Sonora Flame Skimmer} disequilibrium chemistry models across a range of $K_{\rm zz}$ values, using synthetic photometry computed for the full grid and mapped to absolute magnitudes using the corresponding evolutionary model radii. Observational comparisons are drawn from the UltracoolSheet compilation \citep{DupuyLiu2012, DupuyKraus2013, Deacon2014, Liu2016, Best2018, Best2021, Sanghi2023, Schneider2023, best_2025_zenodo}, where subdwarfs are shown as yellow triangles, young objects as blue squares, and field objects as gray circles.

As cloud-free models, the \texttt{Sonora Flame Skimmer} grid does not reproduce the red $J-K$ colors of late L dwarfs prior to the L-T transition, which require cloudy atmospheres \citep{Morley2024}. Additionally, there is little distinction between models with different $K_{\rm zz}$ values in these near-IR bandpasses. The offset and redness of the Y-J tracks go back to a similar source. Our models are systematically brighter in the J band. This can be due to the fact that for late T dwarfs, cloud species like Cr, MnS, Na2S, ZnS, and KCl impact the Y and J band brightness \citep{Morley2012}. Earlier Sonora opacity treatments followed the pressure-broadened alkali framework of \citet{Burrows2000}, while the updated opacity set used in recent Sonora models incorporates newer unified alkali line profiles from \citet{Allard2016, Allard2019}. These profiles are designed to capture the far wings of the Na and K resonance doublets in H$_2$/He-rich atmospheres, where standard Lorentzian or molecular pressure-broadening prescriptions are insufficient. Uncertainties in these alkali line wings may contribute to residual offsets in the $Y$ band and therefore in $Y-J$ colors. Larger differences are also seen in W1 and W2, where the carbon chemistry is dictated by the strength of vertical mixing.

\begin{figure*}
    \centering
    \includegraphics[width=\textwidth]{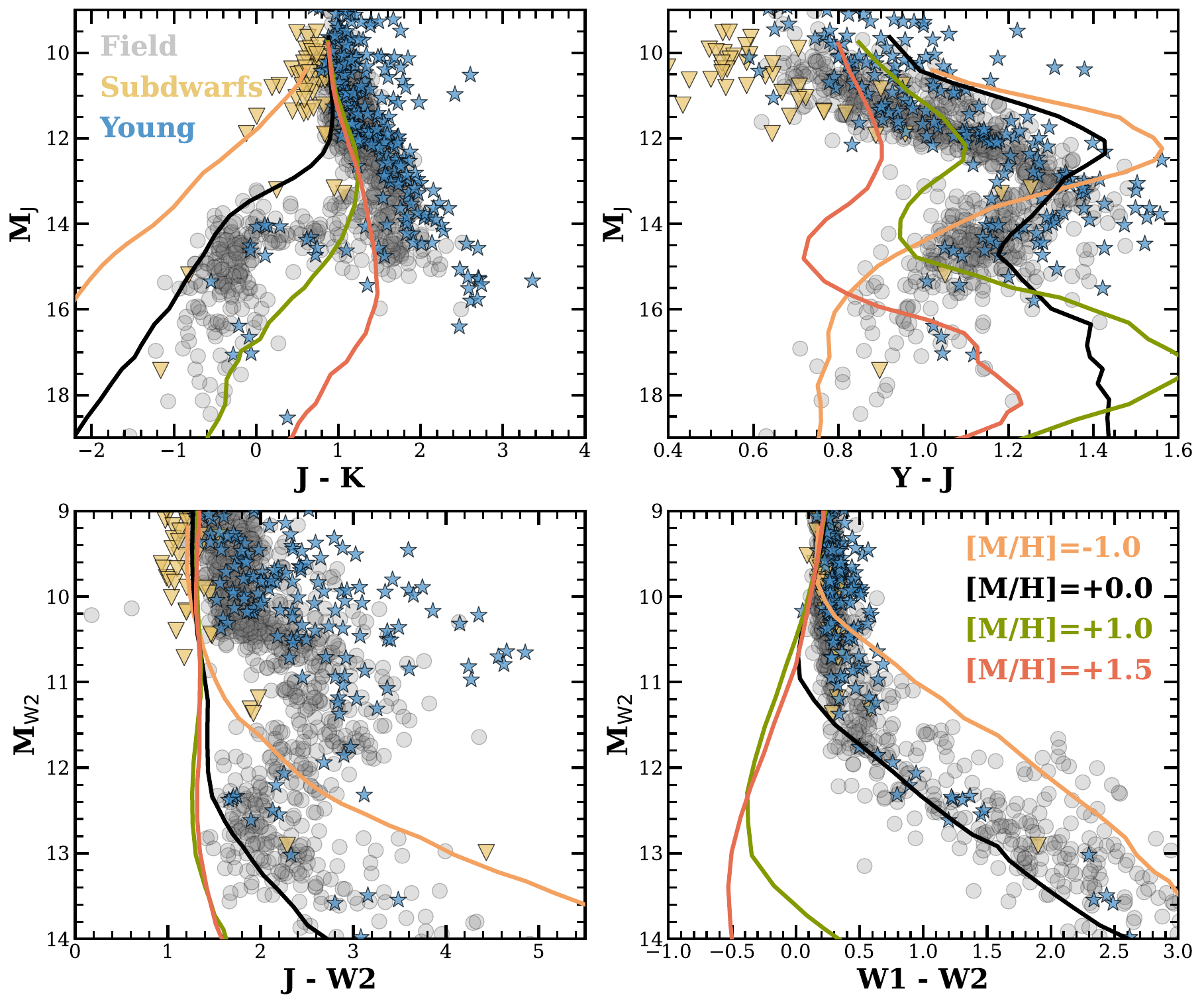}
    \caption{Color–magnitude diagrams using MKO and WISE (W1, W2) filters, with a comparison sample drawn from the UltracoolSheet. \texttt{Sonora Flame Skimmer} disequilibrium tracks are shown for $\log(g)=4.5$, C/O = 0.458, and $K_{\rm zz} = 10^7$, with varying [M/H]. Increasing metallicity shifts the tracks to redder colors and lower fluxes.}
    
    \label{fig:cmdmh}
\end{figure*}

Figure~\ref{fig:cmdmh} shows the CMD with the same objects across the same filters but with the \texttt{Sonora Flame Skimmer} models with varying metallicity. High-metallicity models ([M/H] $\gtrsim +1.0$) show systematically bluer colors in the 4-5~$\mu$m region due to enhanced CO$_2$ and CO absorption. The sub-solar metallicity track exhibits less monotonic behavior in Y-J color due to reduced molecular opacity from H$_2$O and CH$_4$, which weakens the differential absorption between the Y and J bands. In this regime, collision-induced absorption (CIA) from H$_2$ becomes comparatively more important, contributing a broad and relatively featureless opacity source that suppresses both bands more uniformly and prevents the turnover seen at higher metallicities. Meanwhile in W1 - W2, the sequence of metallicities follows the expected trend due to the change in abundances of species like CO$_2$ and CO in these near-IR bandpasses.

\section{Discussion} \label{sec:discussion}

\subsection{Model Convergence} \label{sec:convergence}

While \texttt{Sonora Flame Skimmer} is constructed as a uniform, square grid across all parameters, certain regions of parameter space correspond to extreme evolutionary stages—either very early, near formation, or at ages exceeding that of the universe. As a result, we do not place strong emphasis on irregularities that arise in these regimes, such as combinations of high temperature with low surface gravity or low temperature with high surface gravity. In practice, this can produce repeated entries in the evolutionary tables for a given mass and age across a range of cold temperatures at fixed surface gravity, reflecting the boundaries of the model grid rather than physically distinct solutions. In addition, we encountered convergence difficulties for a subset of high-metallicity models ([M/H] = +1.5 and +2.0) at early L dwarf temperatures. In these cases, the models develop extended convective zones that reach high into the atmosphere, complicating numerical convergence. We note that for the convective regions in the \texttt{PICASO} RCE solver, it follows an adiabatic lapse rate computed assuming a pure H/He mixture with solar abundances, including for the metal-rich atmospheres in this grid. At the highest metallicities explored here, deviations from a non-solar H/He adiabatic lapse rate may alter the thermal structure and locations of the convective zones.

At colder temperatures, particularly across the T–Y dwarf transition, achieving convergence often required initializing the models with slightly warmer thermal profiles. For example, a converged 500 K model may require an initial profile closer to 525 K. This behavior is driven by the presence of detached convective zones, which introduce additional complexity into the thermal structure. To improve convergence in this regime, we also limit the magnitude of temperature adjustments permitted at each iteration. This constraint is critical for resolving the intervening radiative zone and preventing it from being numerically bypassed during iterations. This approach is applied to late T dwarfs and all Y dwarf models. The specific thresholds and implementation of this temperature-limiting scheme in \texttt{PICASO} are described in detail in \citet{Mang2026}. There is ongoing development to improve the radiative-convective zone solver that will further improve and simplify these cases and conditions.

Additionally, for some disequilibrium models colder than 300 K, chemical timescales become sufficiently long that, for high values of $K_{\rm zz}$, vertical mixing dominates deep in the atmosphere. As a result, species quench at very high pressures ($\gtrsim 10^{4}$ bar), often below the maximum pressure boundary of the atmospheric profile. To accurately capture these quenched abundances, we extend the atmospheric profiles downward to $10^{6}$ bar along the adiabat and compute the chemical abundances over this extended region. We then identify the quench points deep in the atmosphere and remove the extended pressure layers, reverting to the original pressure grid while retaining the computed abundances.

\subsection{Clouds} \label{sec:clouds}
Clouds are not included in \texttt{Sonora Flame Skimmer}. For the L dwarfs warmer than $T_{\rm eff} \sim$ 1300 K, iron and silicate clouds should condense in the atmosphere. These cloudy models exist as \texttt{Sonora Diamondback}, although it does not have as wide a metallicity range as \texttt{Sonora Flame Skimmer}. The additional opacity source from clouds warms the atmospheric profile and impacts the evolutionary tracks as seen in \citet{Morley2024}. The colder models in our grid with $T_{\rm eff} \leq$ 450 K are within the regime where water, ammonia, and even methane clouds can begin to condense \citep{Morley2012, Morley2014, faherty2014, leggett2015, skemer2016, Morley2018, Mang2022, Lacy2023, Kuhnle2025}. These clouds are seen in the atmospheres of the gas giants in our own Solar System \citep{Sato1979, Carlson1992, Banfield1998, Guillot2020}. The additional opacity from these clouds will dampen the observed spectral features, especially in the near-IR where there are clear water-ice features \citep{Morley2014, Mang2022, Lacy2023}. The inclusion of clouds will also alter the atmospheric profiles. With the additional opacity source, the cloud acts as a blanket to trap more heat emerging from the deeper atmosphere, which while suppressing the flux in the observable, also shifts the thermal structure of the atmosphere due to radiative feedback heating the deeper atmosphere \citep{Mang2024}. Since the thermal structures are used as the boundary conditions for the evolutionary models, the addition of clouds, as seen in \texttt{Sonora Diamondback} will change the evolution timescales of these substellar objects. The impact of clouds on the evolution of substellar objects is expected to increase with metallicity, as metal-rich atmospheres contain larger abundances of condensable species (e.g., H$_2$O) that can form clouds, enhance atmospheric opacity, and further slow radiative cooling. Future work will include volatile clouds self-consistently within the colder range of Sonora models. 

\subsection{Implications for JWST \& Future Observations} \label{sec:JWST}

\begin{figure}
    \centering
    \includegraphics[width=\columnwidth]{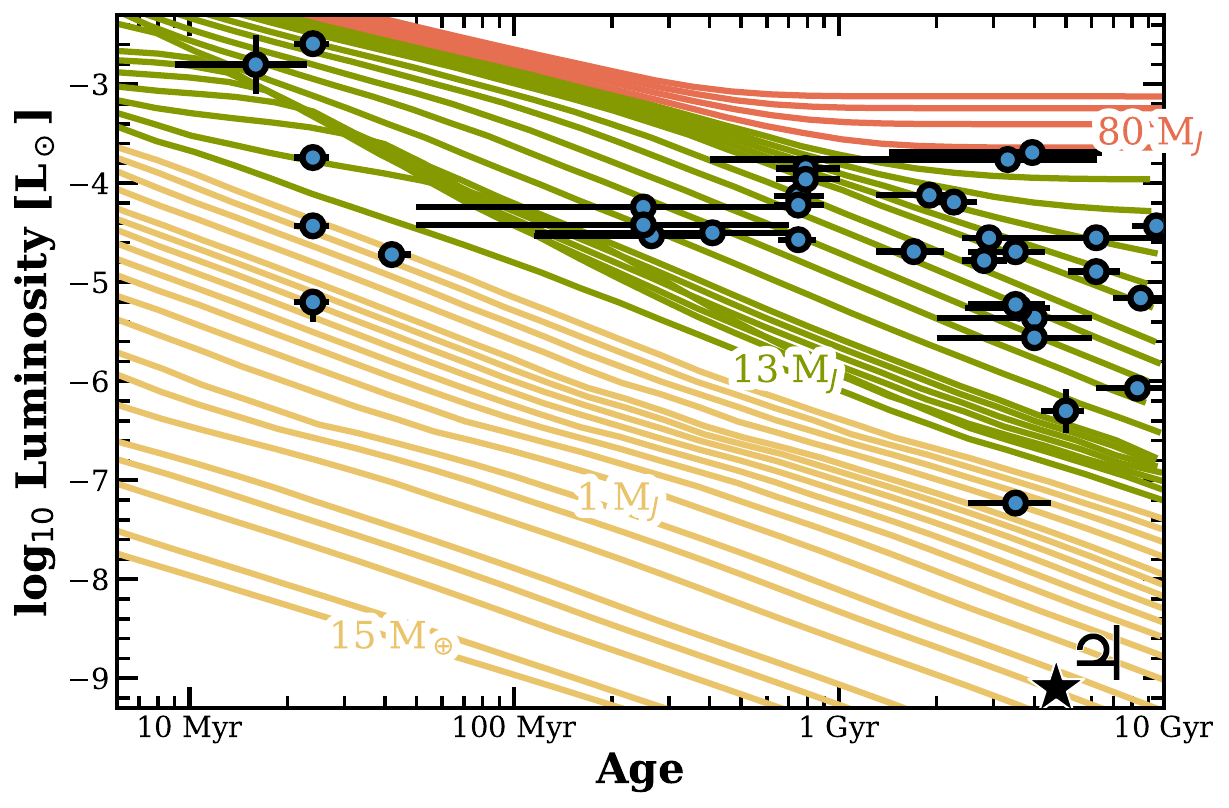}
    \caption{\texttt{Sonora Flame Skimmer} equilibrium evolutionary tracks categorized by mass for planets (yellow), brown dwarfs (green), and stars (red). Directly imaged dynamical benchmark objects compiled in \citet{Li2026}, along with $\epsilon$~Indi~Ab \citep{Sanghi2026epsindi}, are shown as blue circles, with Jupiter shown as a black star.}
    \label{fig:benchmarks}
\end{figure}

Figure~\ref{fig:benchmarks} overplots a sample of dynamical benchmark systems on the \texttt{Sonora Flame Skimmer} evolutionary tracks. The benchmark sample is drawn from the compilation of \citet{Li2026} and includes recent measurements of Eps Indi Ab \citep{Sanghi2026epsindi}. JWST will continue to populate the gaps in the colder, lower-mass object regime across all ages, moving toward true Solar System analogs. The \texttt{Sonora Flame Skimmer} models presented here extend the Sonora framework to sufficiently low temperatures and masses to overlap with this emerging population. As a result, these models enable more robust color-based selection criteria for identifying and prioritizing candidates for follow-up observations in JWST surveys.

\begin{figure}
    \centering
    \includegraphics[width=\columnwidth]{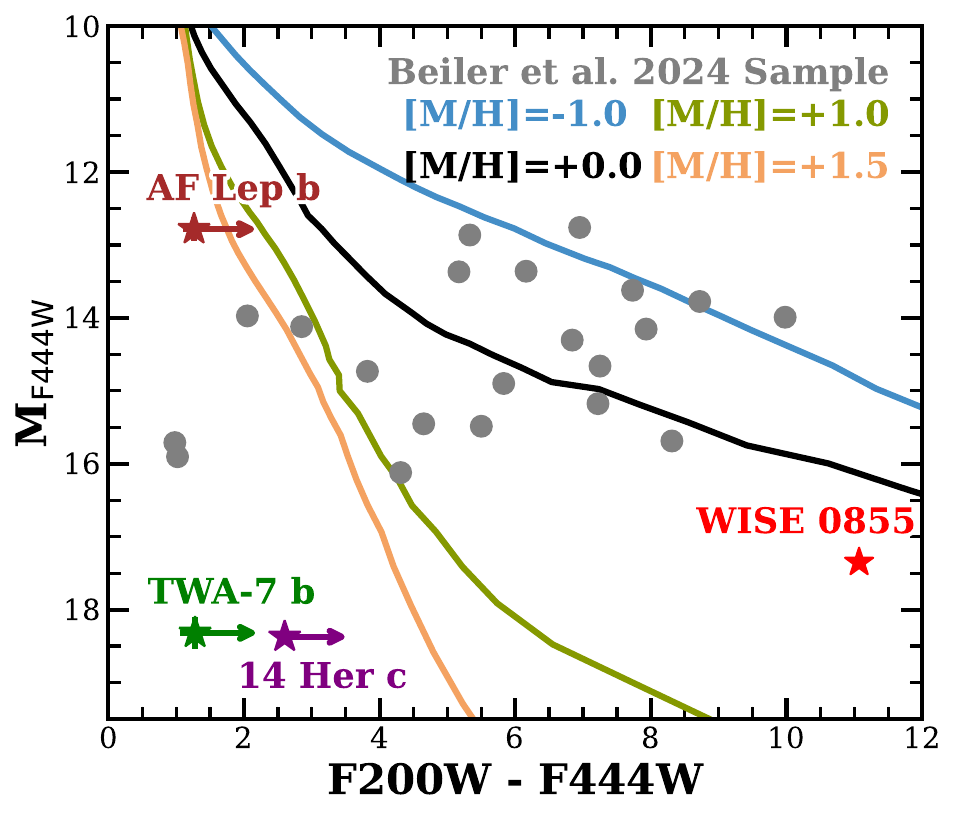}
    \caption{Color–magnitude diagram showing a sample of brown dwarfs from \citet{Beiler2024sample} (gray) alongside \texttt{Sonora Flame Skimmer} tracks at varying metallicities for $\log(g)=4.5$ and C/O = 0.458. WISE 0855 (red) and planetary-mass objects AF~Lep~b (brown), TWA-7~b (green), and 14~Her~c (purple) are shown as stars, with arrows indicating F200W upper limits. The broad spread of these JWST targets demonstrates that metallicity can drive large shifts in observed color at short-wavelength NIRCam observations.}
    \label{fig:jwstcmd}
\end{figure}

Figure \ref{fig:jwstcmd} shows a color-magnitude diagram of the JWST NIRCam F200W and F444W bandpasses, which are commonly used filters for observations of temperate giant planets and other surveys searching for other cold substellar objects (GO 4050, GO 5835, GO 10764 PI: Carter, GO 6005 PI: Biller, GO 6122, 8581 PI: Bowens-Rubin). A sample of brown dwarfs is shown in the gray circles from \citet{Beiler2024sample} along with WISE 0855 (red star) for which we generated synthetic photometry from the JWST NIRSpec/PRISM data from \citet{Luhman2024}. Other early JWST observations of planets AF Lep b \citep{Franson2024}, 14 Her c \citep{BardalezGagliuffi2025}, and TWA-7 b \citep{Crotts2025} are also shown. These planets do not have detections in F200W and thus are upper limits for the color seen in F200W - F444W. Further implications on the sensitivity and yield predictions using \texttt{Sonora Flame Skimmer} are discussed in an upcoming study for these ongoing JWST surveys (Strampelli et al. in prep).

\begin{figure*}
    \centering
    \includegraphics[width=0.48\linewidth]{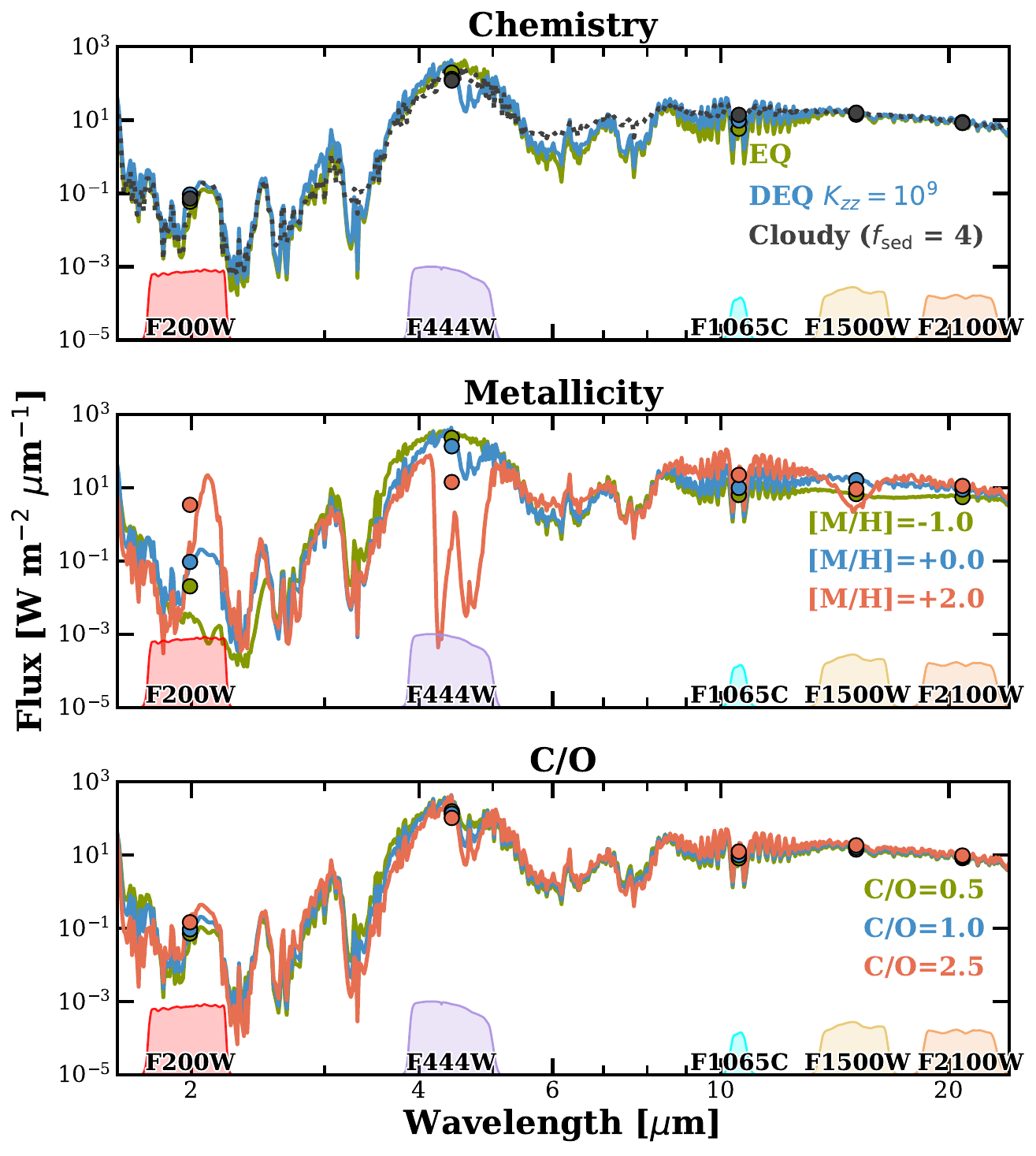}
    \includegraphics[width=0.48\linewidth]{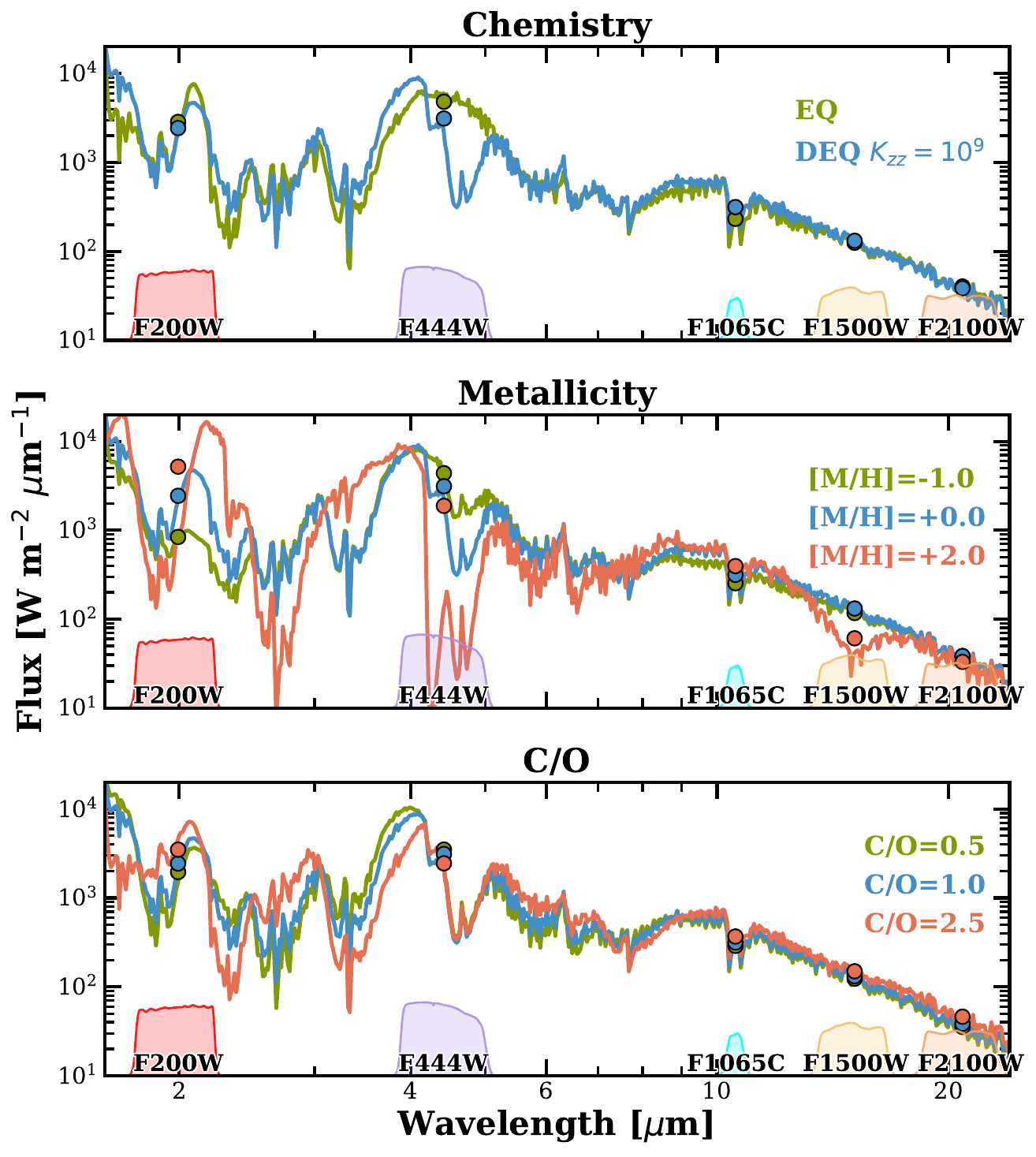}
    \caption{Simulated \texttt{Sonora Flame Skimmer} spectra and synthetic photometry for two representative substellar objects observed with JWST. Left: a cold object with $T_{\rm eff} = 300$~K and log(g) = 4.5. Right: a warmer object with $T_{\rm eff} = 800$~K and log(g) = 4.5. On both sides, the top row compares equilibrium (EQ; green) and disequilibrium (DEQ; blue) chemistry models, the middle row shows the effect of metallicity ([M/H] = $-1.0$, $+0.0$, $+2.0$), and the bottom row shows the effect of the C/O ratio (0.229, 0.458, 1.14). In all rows, the blue spectrum is the same model for reference. Different colored transmission bands for common JWST filters (F200W, F444W, F1065C, F1500W, and F2100W) are overplotted. Metallicity remains the dominating variable in differences observed across the different bands. The effects of the chemistry and C/O ratio, with observable differences driven by the abundance of CO$_2$, become more important with increasing temperatures. For comparison, we also show a 300~K model that includes H$_2$O cloud opacity in the top-left panel (gray), highlighting the potential spectral impact of clouds relative to our cloud-free grid.}
    \label{fig:jwstspec}
\end{figure*}

\begin{figure*}
    \centering
    {\fontsize{12}{14}\selectfont\textbf{Panel A: T$_{\rm eff}$ = 300~K, log(g) = 4.5}}
    \includegraphics[width=\linewidth]{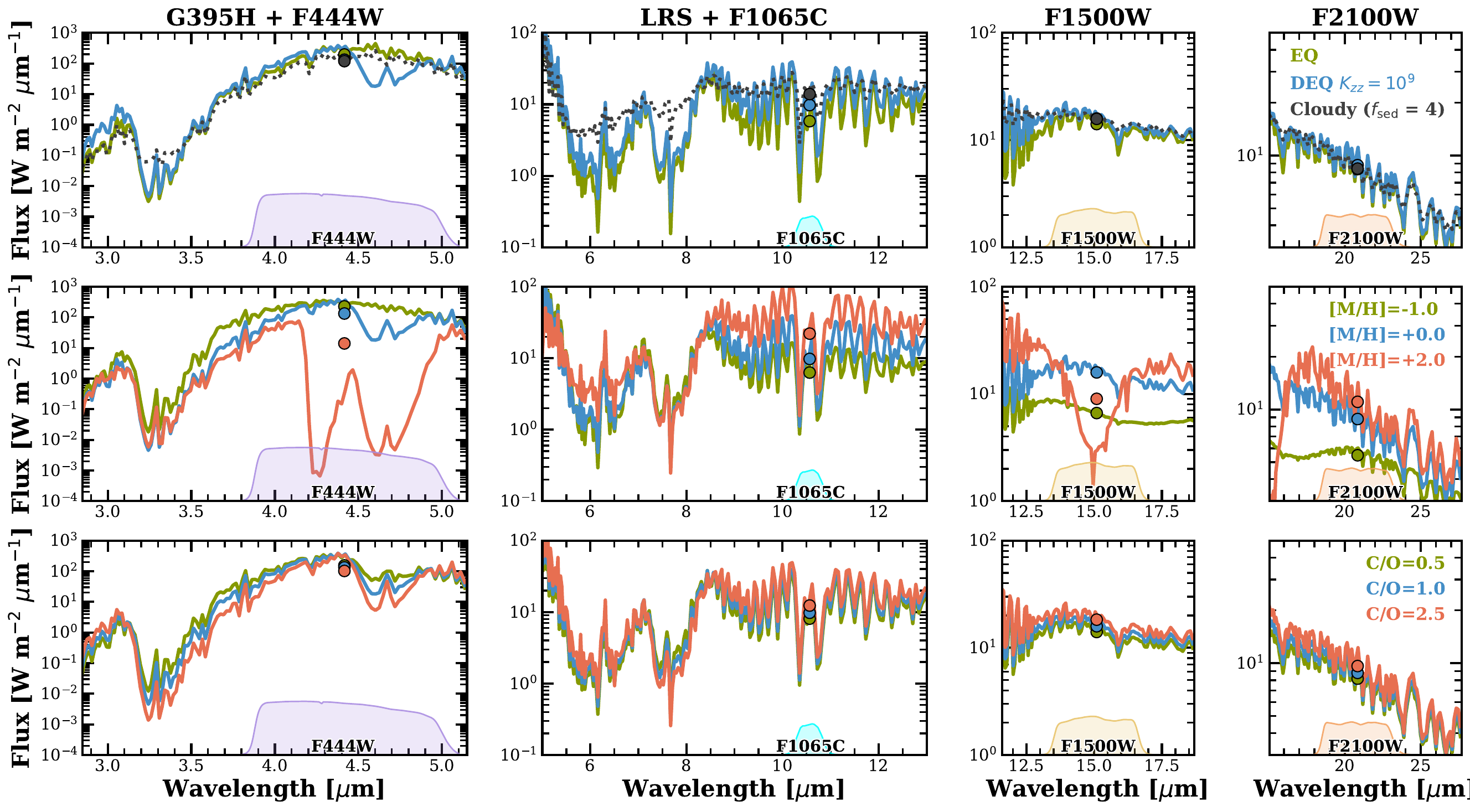}
    {\fontsize{12}{14}\selectfont\textbf{Panel B: T$_{\rm eff}$ = 800~K, log(g) = 4.5}}
    \includegraphics[width=\linewidth]{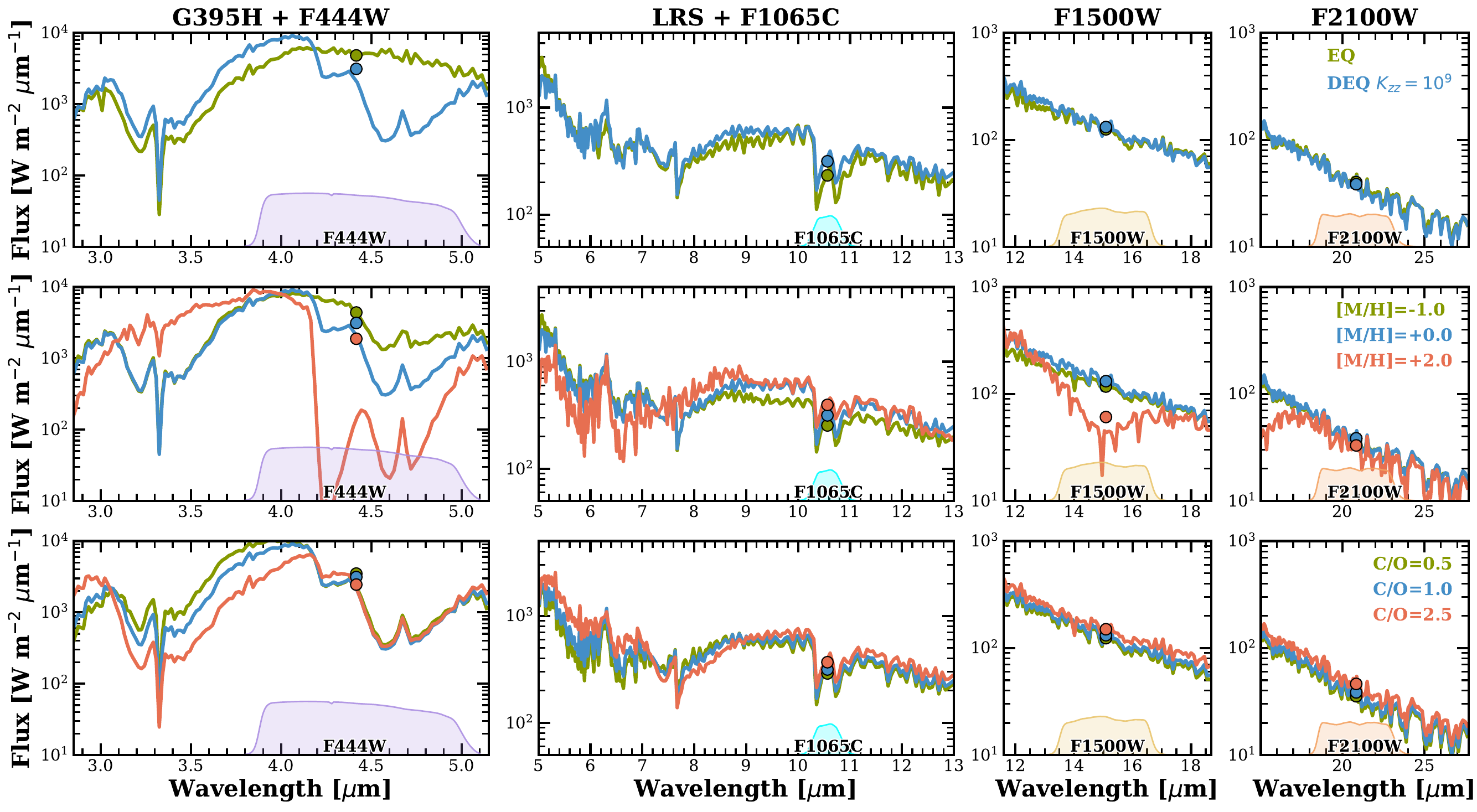}
    \caption{Simulated \texttt{Sonora Flame Skimmer} spectra and synthetic photometry for two representative substellar objects observed with JWST. Top: a cold object with $T_{\rm eff} = 300$~K and log(g) = 4.5. Bottom: a warmer object with $T_{\rm eff} = 800$~K and log(g) = 4.5. For both objects, the top row compares equilibrium (EQ; green) and disequilibrium (DEQ; blue) chemistry models, the middle row shows the effect of metallicity ([M/H] = $-1.0$, $+0.0$, $+2.0$), and the bottom row shows the effect of the C/O ratio (0.229, 0.458, 1.14). Different colored transmission bands for common JWST filters (F444W, F1065C, F1500W, and F2100W) are overplotted. Metallicity remains the dominating variable in differences observed across the different bands. The effects of the chemistry and C/O ratio, with observable differences driven by the abundance of CO$_2$, become more important with increasing temperatures. For comparison, we also show a 300~K model that includes H$_2$O cloud opacity in the top-left panel (gray), highlighting the potential spectral impact of clouds relative to our cloud-free grid.}
    \label{fig:jwstspecmixed}
\end{figure*}

Figures~\ref{fig:jwstspec} and \ref{fig:jwstspecmixed} illustrate the effects of varying chemistry, metallicity, and C/O on the emergent spectra. We show representative models at $T_{\rm eff} = 300$~K and 800~K across the JWST wavelength range relevant for planetary mass companions, along with synthetic photometry in commonly used JWST bandpasses, including NIRCam F200W and F444W and MIRI F1065C, F1500W, and F2100W. The 300~K case represents temperate gas giants such as 14~Her~c, TWA-7~b, and $\epsilon$~Indi~Ab, while the 800~K case is representative of objects like AF~Lep~b. Differences between equilibrium and disequilibrium chemistry are primarily driven by carbon chemistry, most notably through enhanced CO$_2$ abundances, which produce strong variations in bands such as F444W. Metallicity has the largest overall impact, systematically shifting the abundances of dominant species (e.g., H$_2$O, NH$_3$, CO, CO$_2$, and CH$_4$) and producing broad changes across the spectrum. In contrast, variations in C/O ratio introduce more modest differences, with their effects becoming more noticeable at higher temperatures.

A notable trend emerging from recent JWST observations of temperate giant planets is their relative faintness in the F444W band. This suppression may arise from a combination of cloud opacity and molecular absorption, including the CO$_2$ feature near 4.2~$\upmu$m produced by disequilibrium carbon chemistry. \citet{Bowens-Rubin2025} investigated this effect for cold, cloud-bearing atmospheres by comparing JWST sensitivity in F444W and F2100W. They found that variations in cloud coverage can produce brightness differences of up to 10 magnitudes in F444W, while H$_2$O clouds have a negligible impact at F2100W. 

For comparison, we include an illustrative model in Figures~\ref{fig:jwstspec} and \ref{fig:jwstspecmixed} that self-consistently includes H$_2$O cloud opacity in a 300~K atmosphere. The clouds in this model follow the prescription of \citet{Ackerman2001} and adopt a sedimentation efficiency parameter of $f_{\rm sed}=4$. Larger values of $f_{\rm sed}$ correspond to a more cloudy atmosphere, with $f_{\rm sed}=10$ representing an effectively cloud-free atmosphere. Further details on the cloud coupling to the climate model are described in \citet{Mang2026}. The cloud effects in this model are modest, but are expected to become increasingly important at colder temperatures as more H$_2$O condenses into a cloud, becoming more optically thick. Our next generation of Sonora models will incorporate clouds alongside both equilibrium and disequilibrium chemistry to capture these effects self-consistently across the grid.

On the horizon is the launch of the Nancy Grace Roman Space Telescope (NGRST). Its Coronagraph Instrument, while a technology demonstration, may reach down to Saturn-sized planets and deliver images and low-resolution spectra of mature, Jupiter-like exoplanets \citep{Lupu2016, Lacy2019, Morley2014}, potentially revealing more information about the disequilibrium carbon chemistry and water clouds in their atmospheres. Reflected-light observations are highly sensitive to clouds \citep{Sudarsky2000, Sudarsky2005, Lupu2016, Sanghi2026epseri}, meaning that uncertainties in H$_2$O and NH$_3$ cloud formation will directly impact the expected contrast of these objects and thus their detectability.

\section{Conclusions} \label{sec:conclusion}

We present \texttt{Sonora Flame Skimmer}, a new grid of cloud-free 1D atmospheric and evolutionary models that extends the Sonora family to colder temperatures, lower surface gravities, and a broader range of metallicities and C/O ratios. The grid spans $T_{\rm eff} = 50$-2400~K, $\log(g) = 2.0$-5.5 (cgs), metallicities [M/H] = $-1.0$ to $+2.0$ (0.1$\times$ to 100$\times$ solar), and C/O ratios from solar to 2.5$\times$ solar. The models include both chemical equilibrium and disequilibrium chemistry with vertical mixing strengths of $K_{\rm zz} = 10^2$-$10^9$ cm$^2$ s$^{-1}$. These models are designed to interpret the growing population of cold substellar objects observed with JWST and to prepare for future observations of Solar System analogs. All \texttt{Sonora Flame Skimmer} atmospheric structures, chemical abundance profiles, spectra, synthetic photometry, and evolutionary tracks are publicly available on Zenodo for the community to use with the following doi: \href{https://doi.org/10.5281/zenodo.20030439}{10.5281/zenodo.20030439}. Key results from this work include:

\begin{enumerate}
    \item \textbf{Expanded parameter space:} \texttt{Sonora Flame Skimmer} extends atmospheric and evolutionary models down to 50~K and $\log(g)=2$, enabling the study of Neptune-like objects and low-gravity planetary-mass companions for the first time within the Sonora framework.
    
    \item \textbf{Consistent reproduction of previous grids:} The equilibrium and disequilibrium P-T structures closely reproduce \texttt{Sonora Bobcat} and \texttt{Sonora Elf Owl} models, validating the updated \texttt{PICASO}-based framework.
    
    \item \textbf{Updated disequilibrium chemistry:} The inclusion of improved CO$_2$ kinetics increases CO$_2$ abundances and produces corresponding suppression of flux in the 4--5~$\mu$m region. The metallicity-dependent quenching also introduces smaller but measurable changes in the abundances of CO, CH$_4$, H$_2$O, and HCN.

    \item \textbf{Condensation of volatile species without cloud opacity:} In disequilibrium models, we allow H$_2$O, NH$_3$, and CH$_4$ to condense and be depleted from the gas phase once they reach their saturation vapor pressures. This produces weaker absorption features relative to the \texttt{Sonora Elf Owl} models, while still omitting cloud opacity.

    \item \textbf{Updated evolution models:} While \texttt{Sonora Flame Skimmer} largely matches most of the \texttt{Sonora Bobcat} evolutionary tracks, the largest differences are seen at younger, more massive objects as we use the SPHINX models to extend the atmospheric boundary condition. Other differences at lower masses stem from the inclusion of a core in objects $\leq 3~M_{\rm J}$ and updates to the EOS (SCvH to CMS).
    
    \item \textbf{Metallicity-driven evolution:} Higher metallicity atmospheres slow the cooling of substellar objects, leading to higher luminosities and effective temperatures at the same age. At solar metallicity, equilibrium and disequilibrium evolutionary tracks show minimal differences, indicating that atmospheric chemistry primarily affects spectra rather than global evolution. Meanwhile variations in vertical mixing strength ($K_{\rm zz}$) have a negligible impact on bulk evolution. We also find that in the highest metallicity case, the DBMM and HBMM are significantly reduced to 5.39 and 45.03 $M_{\rm J}$, respectively.
\end{enumerate}

With ongoing JWST observations of ultra-cool Y dwarfs and temperate giant planets, \texttt{Sonora Flame Skimmer} provides a cloud-free baseline for interpreting the effects of composition, disequilibrium chemistry, and evolution in the coldest substellar atmospheres. These models serve as a stepping stone toward identifying and characterizing Solar System analogs by providing broad parameter-space coverage that can guide observations. These models will be critical for refining our understanding of ultra-cool substellar objects and guiding future JWST observations. Future work will incorporate clouds self-consistently into cold substellar atmosphere models, connecting the grid with the \texttt{Sonora Diamondback} models.

\begin{acknowledgments}
The authors would like to thank the referee for their helpful comments which improved the manuscript. J.M. acknowledges support from the National Science Foundation Graduate Research Fellowship Program under Grant No. DGE 2137420. S.M. is supported by the Heising-Simons Foundation through the 51 Pegasi b postdoctoral fellowship. J.J.F. acknowledges support from JWST Theory Grant JWST-AR-03245.003-A. M.S.M. acknowledges support from JWST Theory Grant JWST-AR-01977. This material is based on work supported by the National Aeronautics and Space Administration under grant No. 80NSSC24K0958 for the NASA XRP program. Support from program JWST-GO-02507.004-A, JWST-GO-2243, JWST-AR-03245.004-A, JWST-GO-04403.005-A, JWST-GO-02327, JWST-GO-04050.009-A, JWST-AR-01977.004 was provided by NASA through a grant from the Space Telescope Science Institute, which is operated by the Association of Universities for Research in Astronomy, Incorporated, under NASA contract NAS5-26555. This work has benefited from The UltracoolSheet at \href{http://bit.ly/UltracoolSheet}{http://bit.ly/UltracoolSheet}, maintained by Will Best, Trent Dupuy, Michael Liu, Rob Siverd, and Zhoujian Zhang, and developed from compilations by \citet{DupuyLiu2012, DupuyKraus2013, Liu2016, Best2018, Best2021, Sanghi2023, Schneider2023}.
\end{acknowledgments}

\begin{contribution}
J.M. developed the \texttt{PICASO} software capabilities required to generate the grid, produced the atmospheric models, led the manuscript preparation, and generated the final publicly available data products. Y.C. generated the evolutionary models. N.E.B. contributed to the software development of \texttt{PICASO} and generated the high-resolution spectra. N.F.W. implemented the updated CO$_2$ kinetic prescription used in \texttt{Sonora Elf Owl v2}. S.M. provided guidance on the generation of disequilibrium chemistry models within \texttt{PICASO}. C.V.M., J.J.F., and M.S.M. provided guidance on the generation of the grid and interpretation of the results. C.V. provided guidance on the chemical prescriptions adopted in the grid. All authors contributed to writing and editing the manuscript.
\end{contribution}

%

\vspace{5mm}


\software{\texttt{PICASO} \citep{Batalha2019,Mukherjee2023,Mang2026}, \texttt{Jupyter} \citep{kluyver2016jupyter}, \texttt{NumPy} \citep{walt2011numpy}, \texttt{SciPy} \citep{SciPy}, and \texttt{Matplotlib} \citep{matplotlib}. Models from this paper are hosted on Zenodo for public use.}



\appendix

\section{\texttt{Sonora Bobcat} \& \texttt{Sonora Elf Owl} P-T Profile Benchmarks} \label{sec:bobcat_elf_pt}

\begin{figure*}
    \centering
    \includegraphics[width=\textwidth]{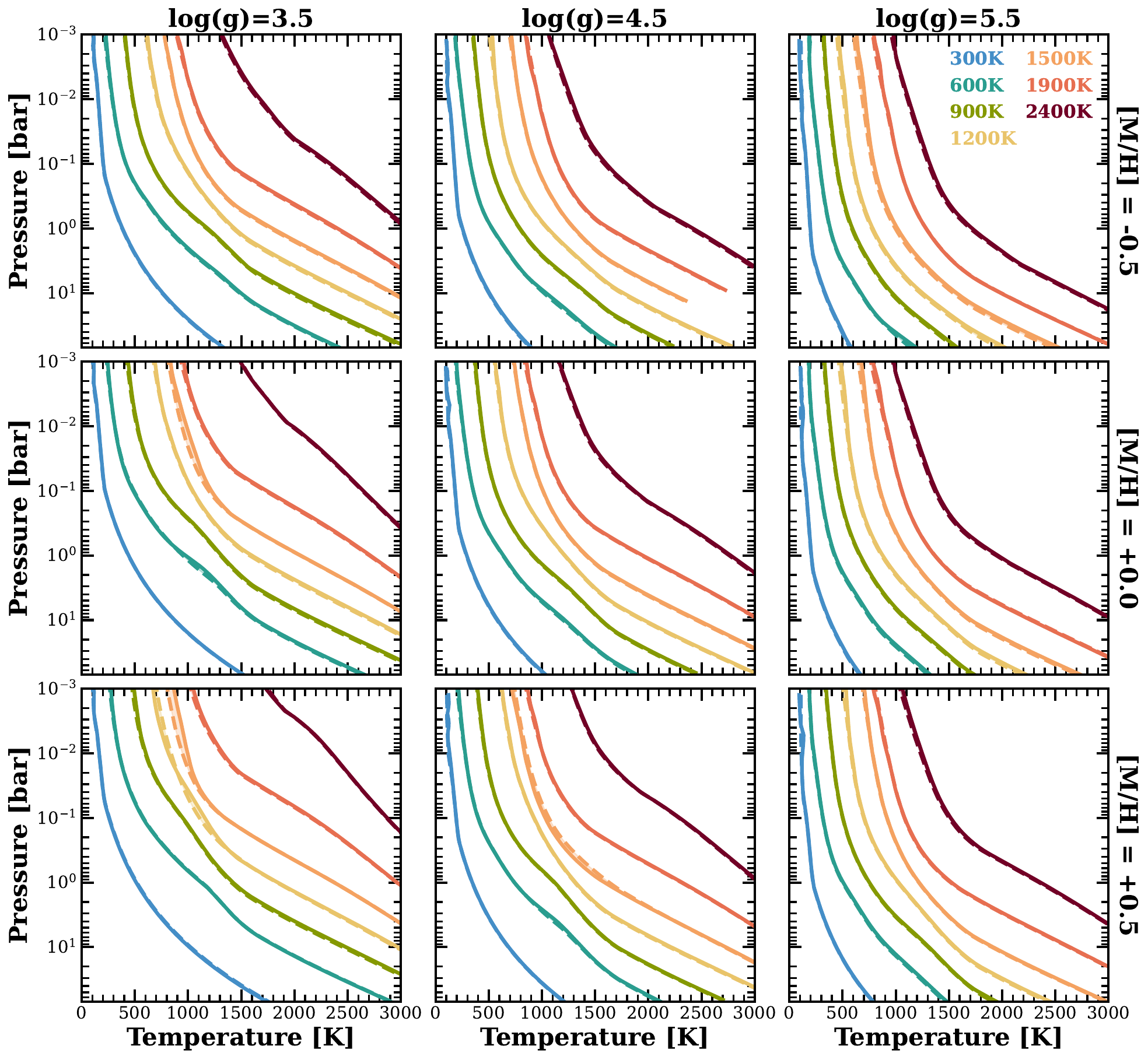}
    \caption{Comparison of pressure–temperature profiles between \texttt{Sonora Flame Skimmer} (solid) and \texttt{Sonora Bobcat} (dashed) at $T_{\rm eff} = 300, 600, 900, 1200, 1500, 1900,$ and $2400$~K. Columns correspond to different surface gravities ($\log(g)=3.5, 4.5, 5.5$), and rows to different metallicities ([M/H] = $-0.5$, $+0.0$, $+0.5$), with all models assuming C/O = 0.458. The close agreement between the two grids across parameter space validates the \texttt{Sonora Flame Skimmer} implementation of equilibrium chemistry and thermal structure.}
    \label{fig:bobcatpt}
\end{figure*}

\begin{figure*}
    \centering
    \includegraphics[width=\textwidth]{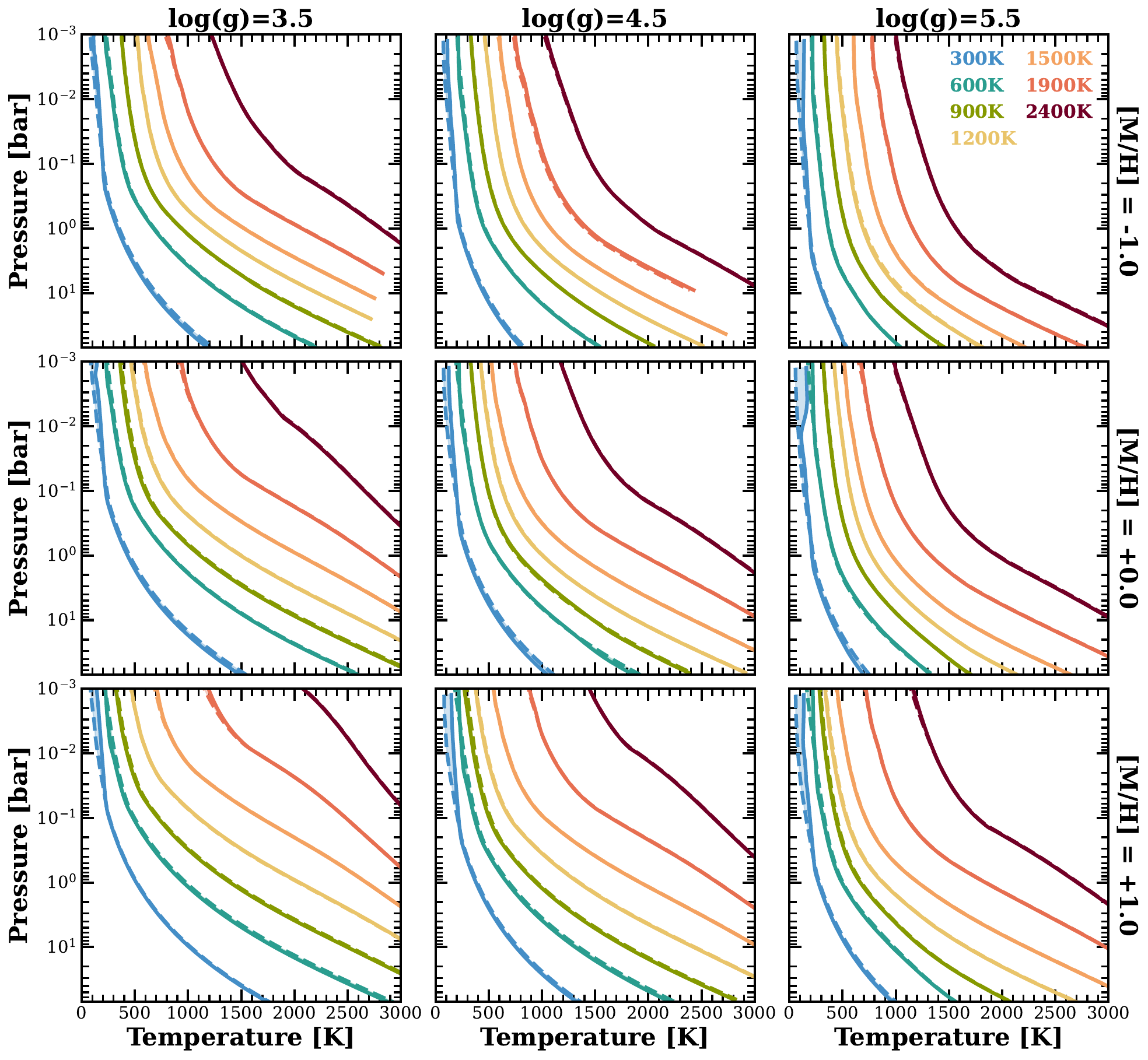}
    \caption{Comparison of pressure–temperature profiles between \texttt{Sonora Flame Skimmer} (solid) and \texttt{Sonora Elf Owl} (dashed) at $T_{\rm eff} = 300, 600, 900, 1200, 1500, 1900,$ and $2400$~K. Columns correspond to different surface gravities ($\log(g)=3.5, 4.5, 5.5$), and rows to different metallicities ([M/H] = $-0.5$, $+0.0$, $+0.5$), with all models assuming C/O = 0.458 and $K_{\rm zz} = 10^9$. Shaded regions highlight differences between the profiles. The close agreement across most of parameter space demonstrates that \texttt{Sonora Flame Skimmer} reproduces the \texttt{Sonora Elf Owl} disequilibrium structures, with deviations emerging primarily at the lowest temperatures where volatile condensation occurs.}
    \label{fig:elfpt}
\end{figure*}

Figure~\ref{fig:bobcatpt} compares equilibrium-chemistry \texttt{Sonora Flame Skimmer} P-T profiles to the corresponding \texttt{Sonora Bobcat} profiles over a wide range of effective temperatures (300--2400~K). Each column shows models at fixed surface gravity (log($g$) = 3.5, 4.5, 5.5), while each row varies metallicity ([M/H] = $-0.5$, $+0.0$, $+0.5$). These metallicities are chosen to enable direct comparisons with the \texttt{Sonora Bobcat} grid, which does not extend to other metallicities. The shaded regions highlight the differences between the two sets of profiles. In most cases, these differences are minimal, indicating that \texttt{Sonora Flame Skimmer} reproduces the \texttt{Sonora Bobcat} P–T structures with high fidelity. Small deviations are present in a few cases (e.g., 1200 and 1500 K at $\log(g) = 3.5$ and [M/H] = $+0.5$), but remain negligible. These differences arise from a small difference in the radiative-convective boundary found for the final converged profiles in these cases while the location of the radiative–convective boundary remains largely unchanged otherwise.

We then show an analogous comparison for disequilibrium chemistry models, where \texttt{Sonora Flame Skimmer} profiles are compared to those from \texttt{Sonora Elf Owl} in Figure \ref{fig:elfpt}. The columns and rows of the figure are identical to Figure \ref{fig:bobcatpt}, and the models shown have $K_{\rm zz} = 10^9$. As in the equilibrium case, \texttt{Sonora Flame Skimmer} closely reproduces the \texttt{Sonora Elf Owl} P–T profiles. While updates to the CO$_2$ chemistry and quenching prescriptions have a significant impact on the synthetic spectra (Section \ref{sec:spectra}), they have a minor effect on the thermal structure. Notable differences emerge primarily at the lowest temperatures where at 300 K, deviations become more apparent in the upper atmosphere, where H$_2$O begins to rain out. This process is included in \texttt{Sonora Flame Skimmer}, whereas \texttt{Sonora Elf Owl} maintains constant quenched abundances above the quench level, leading to these differences.

\section{Fits to SPHINX Models}
\label{sec:SPHINX_fits}
Evolution modeling at very early ages, especially for massive sub-stellar objects, requires a relationship between the interior entropy (representated by the equivalent 10 bar temperature as $T_{10}$) and the internal heat flux ($T_{\rm eff}$, that controls cooling) at the top of the atmosphere for $T_{\rm eff} > 2400$ K, i.e. above the boundary conditions provided by our atmospheric models. Linearly extrapolating or clipping this relationship generally leads to poorer predictions compared to full modeling beyond 2400 K \citep{Davis2025}. We use the SPHINX models presented in \citep{Davis2025} to fit a simple model to $T_{10}$ as a function of $T_{\rm eff}$, log$g$, and log of the metallicity $Z$. This enables us to extend our atmospheric boundary conditions beyond $T_{\rm eff} = 2400$ K as well as to use the SPHINX models beyond their log$g$ ($3 - 5.5$) and metallicity ($-0.5 - +0.5$) limits for the \texttt{Sonora Flame Skimmer} models. After some experimentation, we found the following expressions captures the behavior of the SPHINX models well
\begin{equation}
{\rm log}T_{10} =
\begin{cases}
    [a~({\rm log}T_{\rm eff} - 3.3) +& \\
     b~({\rm log}g - 3) + f~{\rm log}Z + c], & {\rm log}T_{10} < d \\
    [e~({\rm log}T_{\rm eff} - 3.3) + & \\
    b~({\rm log}g - 3) + f~{\rm log}Z + c'], & {\rm log}T_{10} >d, 
\end{cases}
\end{equation}
with a total of 6 free parameters ($c'$ is determined by equating the two expressions at log$T_{10}$). We use the \texttt{scipy} non-linear least squares function to find the optimum values of the parameters: $a$ = 0.9399, $b$ = -0.0833, $c$ = 3.619, $d$ = 3.669, $e$ = 2.962, $f$ = 0.0601. The sum of the squares of the residuals is merely 0.185 and more than 95\% of the residuals are smaller than $\pm 0.05$. There is a dramatic change in the slope (compare $a$ and $e$) of the relation between log$T_{10}$ and log$T_{\rm eff}$ at a specific value of log$T_{10} = d$ across all the models. For log$T_{10} > d$, the SPHINX models show that log$T_{10}$ rises very rapidly with log$T_{\rm eff}$, which results in objects ending up on Hayashi tracks at very early ages. We stitch this relation between log$T_{10}$ and log$T_{\rm eff}$ to the one provided by the \texttt{Sonora Flame Skimmer} models by removing any offsets between them at log$T_{\rm eff} = 2400$ K.

\section{Other Evolution Model Comparisons} \label{sec:sm08_appendix}

\begin{figure*}
    \centering
    \includegraphics[width=0.48\linewidth]{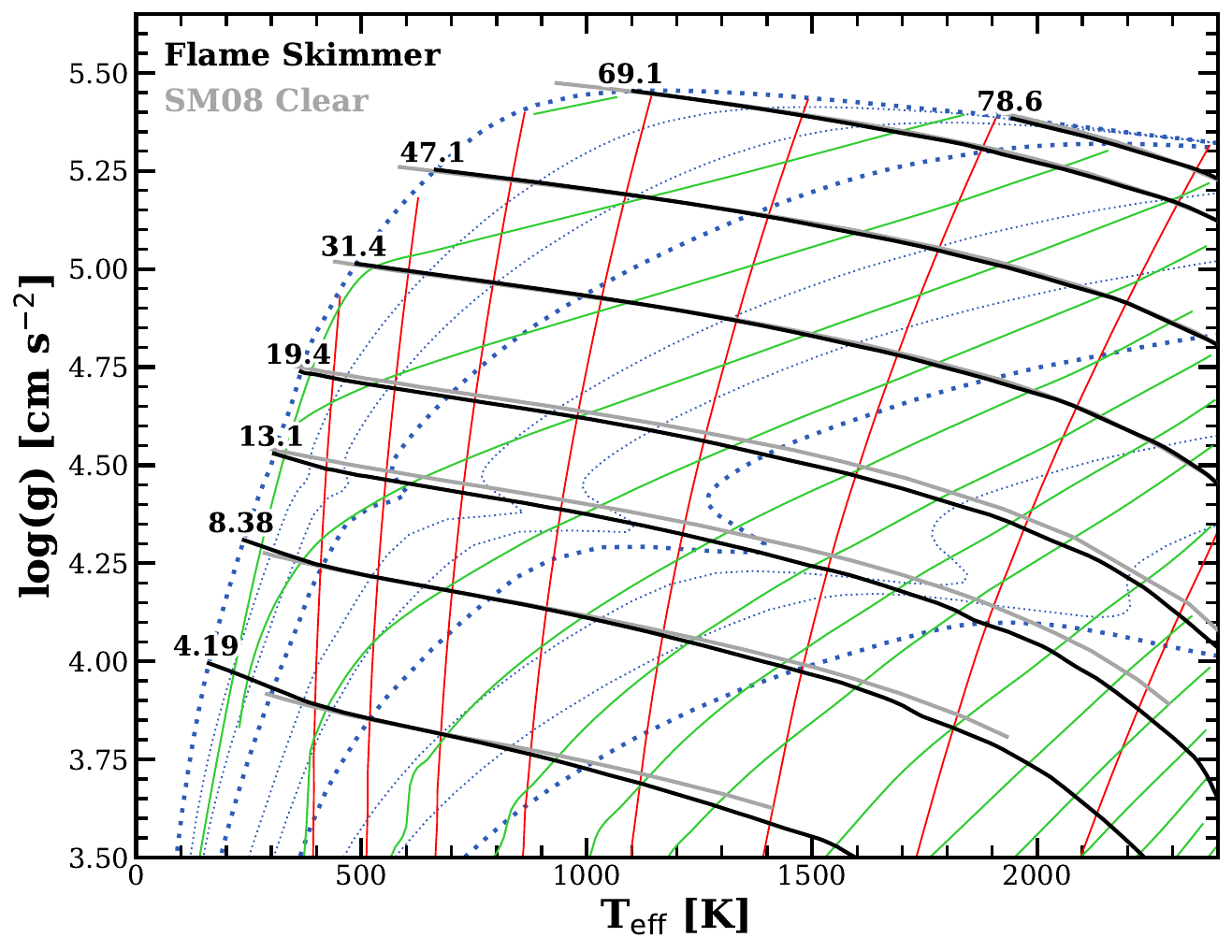}
    \includegraphics[width=0.48\linewidth]{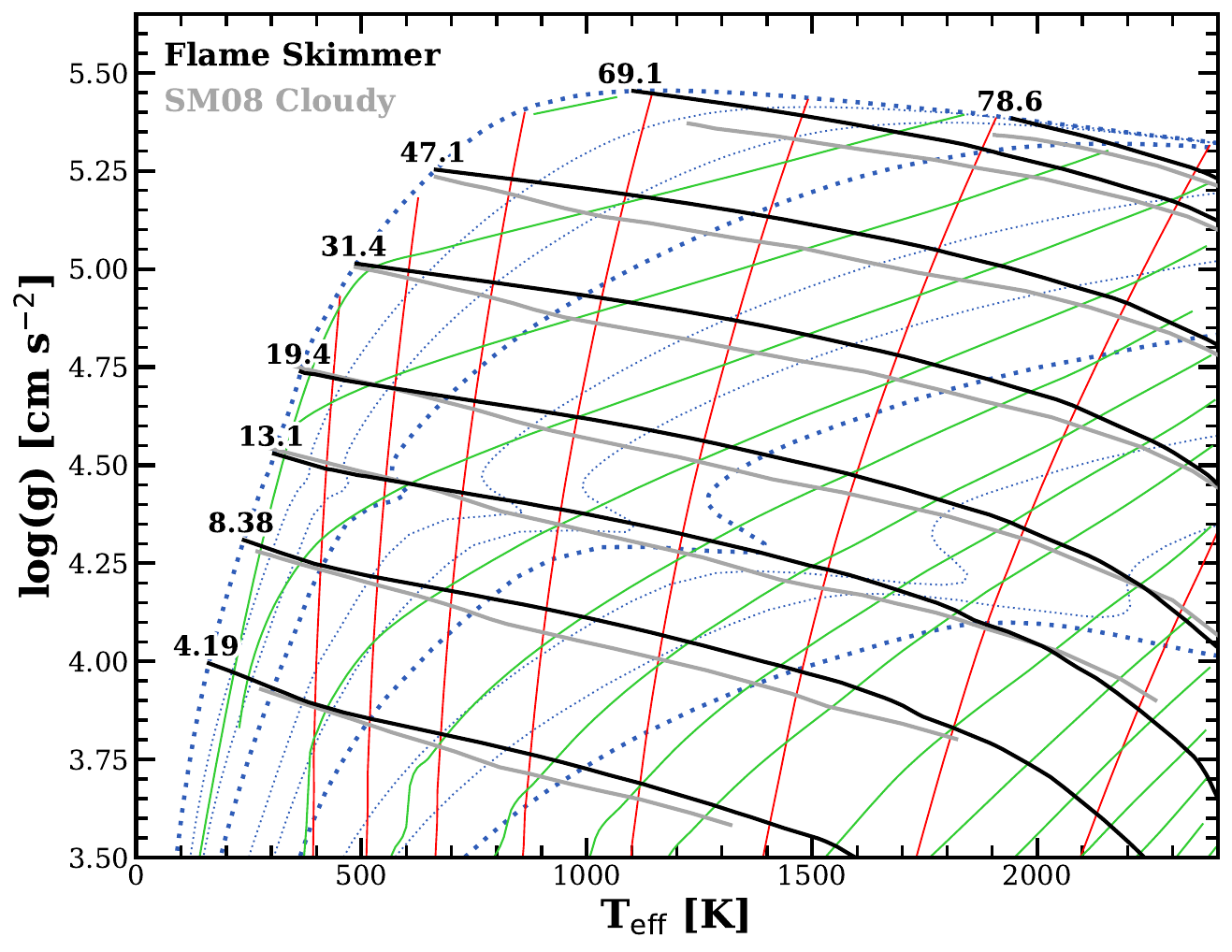}
    \caption{Comparison of evolutionary tracks in $T_{\rm eff}$ and $\log(g)$ space between \texttt{Sonora Flame Skimmer} (black) and \citet{Saumon2008} (gray) models for solar metallicity atmospheres labeled by mass in Jupiter masses. Left: cloud-free (SM08 clear) models. Right: cloudy models with $f_{\rm sed} = 2$ (SM08 Cloudy). Isochrones (blue dotted lines) span 10~Myr to 10~Gyr, with thicker lines marking 0.01, 0.1, 1, and 10~Gyr. Nearly vertical red lines indicate constant luminosity from $\log(L/L_\odot) = -3$ to $-6.5$ in steps of 0.5, and green lines show constant radii from 0.08 to 0.28~$R_\odot$. The \texttt{Sonora Flame Skimmer} models reproduce the overall structure of the SM08 grids in both cloud-free and cloudy cases, with differences primarily arising from updated atmospheric boundary conditions, and more pronounced deviations in the cloudy models due to the additional opacity from cloud formation.}
    \label{fig:sm08_appendix}
\end{figure*}

Figure~\ref{fig:sm08_appendix} compares the evolution of substellar objects in $T_{\rm eff}$–$\log(g)$ space between the \texttt{Sonora Flame Skimmer} models and the \citet{Saumon2008} grids for both cloud-free and cloudy atmospheres. In the cloud-free case (left panel), the two model sets show excellent agreement in higher mass objects but begin to deviate at intermediate mass objects as seen starting after the 19.4 $M_{\rm J}$ mass track. The \citet{Saumon2008} mass track in these intermediate temperatures shows the impact of the inclusion of the new EOS in \texttt{Sonora Flame Skimmer}.

In the cloudy case (right panel), the behavior of the mass tracks matches the expectation as seen in the comparison between cloud-free and cloudy models in \citet{Saumon2008}.  Differences between \texttt{Sonora Flame Skimmer} and \citet{Saumon2008} are more noticeable, particularly at intermediate temperatures where cloud opacity is most influential. These offsets arise from the inclusion of clouds, which alter the atmospheric boundary condition and therefore the cooling rate. This difference shifts the evolutionary tracks toward lower effective temperatures and slightly lower surface gravities at early times, reflecting the reduced radiative efficiency and slower cooling in cloudy atmospheres. 


\newpage
\bibliography{sample63}
\bibliographystyle{aasjournal}



\end{document}